\documentclass[]{amsart}                %-- eventualmente [12pt]
\usepackage{a4wide,amsmath,amssymb,amsfonts,epsfig}  %,filams, colour} %graphicx}
\usepackage{enumerate}                  %--- per personalizzare enumerate
\usepackage{comment}
\usepackage[all]{xy}
\newcommand{\virg}[1]{``#1''}
\def\til{~}
\theoremstyle{plain}
\theoremstyle{definition}
\theoremstyle{remark}
\newtheorem{remark}{Remark}
\newcommand{\parto}[1]{\left(#1\right)}
\newcommand{\parg}[1]{\left\{#1\right\}}

\newcommand{\IN}{{\mathbb N}} % natural
\newcommand{\IR}{{\mathbb R}} % real
\newcommand{\cA}{{\mathcal A}}
\newcommand{\cV}{{\mathcal V}}
\newcommand{\Ov}[1]{\strut\overline{#1}}
\begin{document}
%-------------------------------------
\title[LNG Experiment: Irreversible or Reversible Or Gate?]{A discussion about  LNG Experiment:\\ Irreversible or Reversible Generation of the \textsc{Or} Logic Gate?}
%-------------------------------------
\author[G.~Cattaneo]{Gianpiero Cattaneo}
%---
\address{(Retaired from) Dipartimento di Informatica, Sistemistica e Comunicazione,
         Universit\`a di Milano--Bicocca,
         Viale Sarca 336--U14, I--20126 Milano (Italy)
         }
\email{cattang@live.it}
%-----------------------------
\author[R.~Leporini]{Roberto Leporini}
%---
\address{Dipartimento di Ingegneria Gestionale, dell'Informazione e della Produzione
         Universit\`a di Bergamo,
         Viale Marconi 5, I--24044 Dalmine (Italy)
         }
\email{roberto.leporini@unibg.it}
%-------------
%\author[G.~Vizzari]{Giuseppe Vizzari}
%---
%\address{Dipartimento di Informatica, Sistemistica e Comunicazione,
%         Universit\`a di Milano--Bicocca,
%         Viale Sarca 336--U14, I--20126 Milano (Italy)
%         }
%\email{vizz@disco.unimb.it}
%-------------------------------------
%\thanks{This <The author's> work has been supported by
%   MIUR$\backslash$PRIN project ``Mathematical aspects and forthcoming applications of automata and formal languages''}
%-------------------------------------
%\date{\today}
%---------------------------------------
\keywords{Logic gates, reversible gates, irreversible gates, Landauer principle, LNG device, CL reversible gate, Toffoli reversible gate}
%-------------------------------------
\begin{abstract}
In a recent paper M.\til Lopez-Suarez, I.\til Neri, and L.\til Gammaitoni (LNG) present a concrete realization of the Boolean \textsc{Or} irreversible gate, but contrary to the standard Landauer principle, with an arbitrary small dissipation of energy. A Popperian good falsification!

In this paper we discuss a theoretical description of the LNG device which is indeed a 3in/3out self--reversible realization of the involved \textsc{Or} gate, satisfying in this way the Landauer principle of no dispersion of energy, contrary to the LNG conclusions.

The different point of view is due to a different interpretation of the two outputs corresponding to the inputs 10 and 01, which are considered by LNG indistinguishable so producing a non reversible realization of the standard 2in/1out gate.
On the contrary, always considering these two outputs indistinguishable, by a suitable normalization function of the cantilever angles, the experimental results obtained by the LNG device coincide with the \textsc{Or} connective obtained from the third output of the self-reversible 3in/3out CL gate by the \virg{Inputs-Ancilla$\to$Garbage-Output} procedure. Thus, by the self-reversibility this realization is without dissipation of energy according to the Landauer principle.
Furthermore, using the self-reversible  Toffoli gate it is possible to obtain from the LNG device the realization of the connective \textsc{And} adopting another normalization function on the cantilever angles.

Finally, by other suitable normalization procedures on cantilever angles it is possible to obtain also the other logic  \textsc{Nor} and \textsc{Nand} connectives, and in a more sophisticated way the \textsc{Xor} and \textsc{NXor} connectives in a self-reversible way. All this leads to introduce a \emph{universal logic machine} consisting of the \virg{LNG device plus a memory} containing all the necessary angle normalization functions to produce in a self-reversible way, by choosing one of these latter, the logic connectives now listed.

%This is also motivated by some experimental results, but whose almost identical behaviour is only a consequence of the %lake of "sensibility" at the nanometer dimensions of the used instrument, due to the present technology.
\end{abstract}
%-------------------------------------
\maketitle
%-------------------------------------
\section{Introduction}
%-------------------------------------
This paper discusses a recent result obtained by M.\til L\'opez-Su\'arez, I.\til Neri, and L.\til Gammaitoni (LNG) in \cite{LNG16} regarding the link between irreversibility of some logic gates and energy dissipation due to information loss. Quoting LNG from their paper \cite{LNG16} \virg{Popular gates like \textsc{And}, \textsc{Or} and \textsc{Xor}, processing two logic inputs and yielding one logic output, are often addressed as irreversible logic gates, where the sole knowledge of the output logic value is not sufficient to infer the logic value of the two inputs. Such gates are usually believed to be bounded to dissipate a finite minimum amount of energy determined by the input--output information difference.}

From this point of view, \virg{a way to understand irreversibility is to think of it in terms of information erasure. If a logic gate is irreversible, then some of the information input to the gate is lost irretrievably when the gate operates -- that is, some of the information has been erased by the gate. Conversely, in a reversible computation, no information is ever erased, because the input can always be recovered from the output. Thus, saying that a computation is reversible is equivalent to saying that no information is erased during the computation} \cite[pag.\til 153]{NC00}.

The connection between energy consumption and irreversibility is provided by the so--called \emph{Landauer's principle} which can be formulated in two forms:
\begin{description}
\item[Landauer's principle (first form)]
If a computer erases a single bit of classical information, the amount of energy dissipated into the environment is \emph{at least} $k_B T\ ln 2$, where $k_B$ is the \emph{Boltzmann's constant}, and $T$ is the absolute temperature of the environment of the computer (typically in the form of waste heat).
\end{description}
To this form of Landauer's principle an alternative formulation can be given, according to the laws of thermodynamics, not in terms of energy dissipation, but rather in terms of entropy:
\begin{description}
\item[Landauer's principle (second form)]
If a computer erases a single bit of information, the entropy of the environment increases by \emph{at least} $k_B\ ln 2$, where $k_B$ is the Boltzmann constant.
\end{description}
The interesting result of LNG paper is that they claim of presenting \virg{an experiment where  an \textsc{Or} logic gate, realized with a micro--electromechanical cantilever, is operated with energy well below the expected limit [i.e., $k_B T ln 2$], provided the operation is slow enough and frictional phenomena are properly addressed.} This, if true, is a really interesting falsification of the above formulations of Landauer's principle, notable of a great interest from the scientific community interested on this kind of arguments.

In the present paper we take into account this experimental device giving a possible interpretation/description of it as a \emph{reversible} 3in/3out logic gate, and so in agreement with the Landauer's principle (or better its negation involving reversibility and no erasure of information energy in the environment) operating with energy below the expected limit. The 2in/1out \textsc{Or} gate can be recovered fixing one input as ancilla set to the bit 0, and considering two of the outputs as garbage and the remaining output as producing the expected \textsc{Or}. Furthermore, since our gate is not only reversible, but also self--reversible, the serial cascade of two of them produces the identity gate which furnish as global output just the same input, without no real dissipation of information.
%---------------------------------------------------------------
\section{The LNG realization of the OR Logical Gate}
\label{sec:LNG-Or}
%-------------------------------------------------
The device constructed by LNG and described in \cite{LNG16} \virg{consists of a logic switch made with a Si$_3$N$_4$ elastic cantilever $L$ that can be bent by applying electrostatic forces with two electrical probes $P_1$ and $P_2$ closed to the cantilever tip.}

%----
\begin{figure}[ht]
\begin{center}
   \includegraphics[width=8cm]{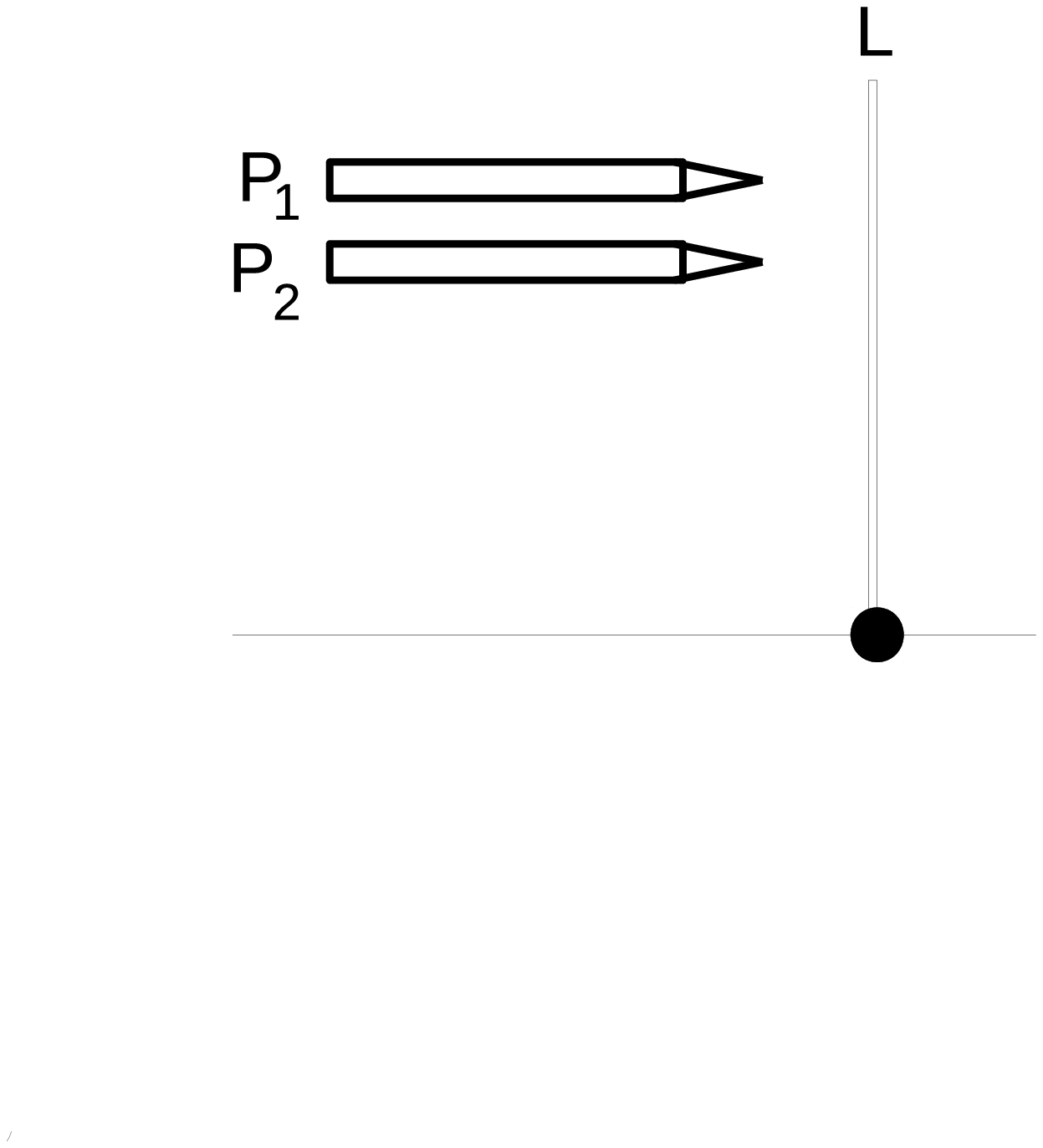}
\end{center}
   \vspace{-1.8cm}
      \caption{Schematic experimental situation consisting of the two electrical probes $P_1$ and $P_2$ which can act by electrostatic forces on the cantilever $L$.}
      \label{fig:LNG}
\end{figure}
%---

Under the initial condition of the cantilever $L$ in the vertical position we can have the following two experimental transitions of the physical system \virg{Probes+Cantilever}:
\begin{enumerate}[(Ex1)]
\item
If no electrode voltage $V$ is applied to the two probes, the cantilever remains in the vertical position.
   %To this vertical position it is associated the logic state 0 of the cantilever.
\item
If on the contrary an electrode voltage $V\neq 0$ is applied to at least one probe, the position of the cantilever is changed as consequence of the electrostatic force.
   %, and in this case we can associate the logic state 1 to the cantilever.
\end{enumerate}

\begin{figure}[h]
\begin{center}
   \includegraphics[width=9cm]{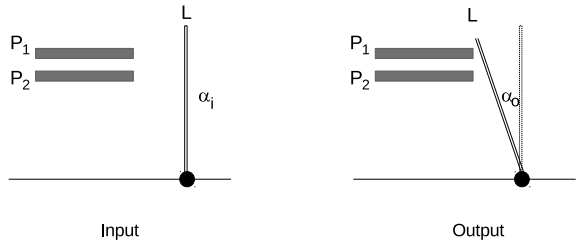}
\end{center}
   \vspace{-1.0cm}
      \caption{Schematic description of the experimental LNG-device where at least one of the probes ($P_1$, $P_2$) is active and the initial cantilever position is vertical ($\alpha_i=0$). Both probes do not change their state and the cantilever $L$ undergoes a deflection of an output angle $\alpha_o\neq 0$.}
      \label{fig:in-out}
\end{figure}
\newpage

Let us stress the following interpretation assumed by the authors of \cite{LNG16} which is the main argument of our analysis:
\begin{enumerate}
\item[(LNG)]
The input of the logic gate is associated with the voltages $V(P_1)$ and $V(P_2)$ of the respective electrical probes $P_1$ and $P_2$. The position of the cantilever tip, measured by its deviation angle $\alpha_o$ with respect to the vertical position, encodes the output of the logic gate. In a first approximation, this corresponds to a 2in/1out gate $G_{LNG}$ formalized by the correspondence:
\begin{equation*}
(V(P_1),V(P_2))\xrightarrow{G_{LNG}}\alpha_o
\end{equation*}
But in the quoted paper it is assumed a drastic convention of associating with the voltages $V(P_i)$ of the probe $P_i$ their \virg{normalized} value $[V(P_i)]=1$ iff $V(P_i)\neq 0$ and $[V(P_i)]= 0$ otherwise, i.e., iff $V(P_i)=0$, corresponding to the logic truth value 1 if the probe is on ($V(P_i)\neq 0$) and the truth value 0 if the probe is off ($V(P_i)=0$).
\\
Similarly for the deviation angle we put $[\alpha]=1$ iff $\alpha\neq 0$, and $[\alpha]=0$ otherwise. With these conventions the authors consider their device as a 2in/1out Boolean gate $G_{LNG}:\{0,1\}^2 \to \{0,1\}$ defined by the correspondence
\begin{equation}
\parto{[V(P_1)],[V(P_2)]}\xrightarrow{G_{LNG}}[\alpha_o]
\end{equation}
\end{enumerate}
As consequence of all these conventions, i.e., interpreting absence or presence of electric voltage on the probes as logic values 0 and 1, respectively, this device realizes the \textsc{Or} logic gate according to the following table which collects all the above remarks:

%\begin{equation}
\begin{table}[h]
      \begin{tabular}{|c|c||c||c|}
         \hline
         $[V(P_1)]$ & $[V(P_2)]$  & $\xrightarrow{LNG}$ &
                 $[\alpha_o]$  \\
         \hline\hline
         \mbox{\rule[0cm]{0cm}{2.5ex}
         $0$}& $0$  & & $0$ \\
         $0$ & $1$  & & $1$ \\
         $1$ & $0$  & & $1$ \\
         \mbox{\rule[0cm]{0cm}{1.5ex}$1$}
             & $1$  & & $1$ \\
         \hline
      \end{tabular}
\vspace{0.3cm}
\caption{Normalized table of the LNG behavior under LNG assumptions}
\label{tbl:LNG-g}
\end{table}
%\end{equation}

 Our position about the LNG realization of the \textsc{Or} gate can be exposed in the following considerations:
 \begin{enumerate}
 \item[(CL)]
 In the transition depicted by Fig.\til \ref{fig:in-out} the input state is applied to the physical system \virg{Probes+Cantilever}, but in order to describe the output state one  must take into account that the device continue to be the whole physical system \virg{Probes+Cantilever}.
 \\
 Therefore, in line of principle, there is no contra-indication to set the initial position of the cantilever in any possible angle $\alpha_i$, of course considering also the particular case $\alpha_i=0$, the input configuration must take into account not only the Boolean pair $[V(P_1)], [V(P_2)]$, but also the Boolean value $[\alpha_i]$.
 \\
 But, since during the transformation the physical device continues to be \virg{Probes+Canti\-le\-ver}, in order to detect the generated output without erasing information about the potentials $V(P_1)$ and $V(P_2)$, which produce the output angle of the cantilever $\alpha_o$, the real output of the device is the configuration $[V(P_1)], [V(P_2)], [\alpha_o]$.
 \\
 This leads to the following table describing the physical transition different from the previous one relative to the input cantilever angle $\alpha_i=0$:

%\begin{equation}
\begin{table}[h]
      \begin{tabular}{|c|c|c||c||c|c|c|}
         \hline
         $[V(P_1)]$ & $[V(P_2)]$ & $[\alpha_i]$ & $\xrightarrow{CL}$ &
                 $[V(P_1)]$ & $[V(P_2)]$ & $[\alpha_o]$ \\
         \hline\hline
         \mbox{\rule[0cm]{0cm}{2.5ex}
         $0$}& $0$ & $0$ & & $0$ & $0$ & $0$\\
         $0$ & $1$ & $0$ & & $0$ & $1$ & $1$ \\
         $1$ & $0$ & $0$ & & $1$ & $0$ & $1$ \\
         \mbox{\rule[0cm]{0cm}{1.5ex}$1$}
             & $1$ & $0$ & & $1$ & $1$ & $1$\\
         \hline
      \end{tabular}
\vspace{0.3cm}
\caption{Normalized table of the LNG behavior under CL assumptions}
\label{tbl:CL-g}
\end{table}
%\end{equation}
 %---
 \end{enumerate}

\begin{figure}[h]
\begin{center}
   \includegraphics[width=8cm]{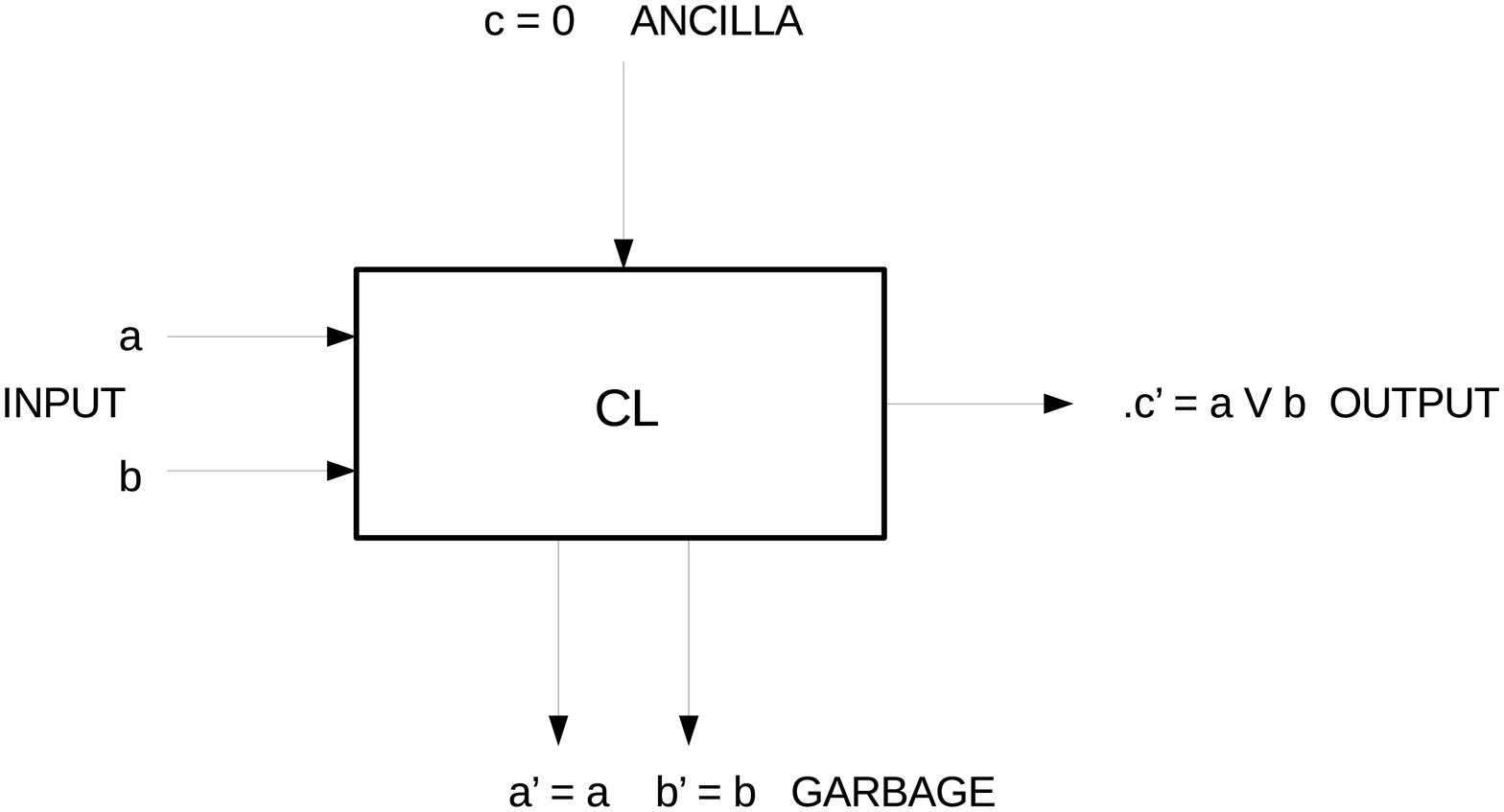}
\end{center}
   \vspace{-1.0cm}
      \caption{Toffoli's box representation of the \textsc{Or} realization according to the CL position. Note that the official Toffoli's terminology is the following: input=argument, output=result, ancilla=source, garbage=sink}
      \label{fig:OR-CL}
\end{figure}

 Of course, this corresponds to a partial reversible non conservative 3in/3out gate, whose complete formulation will be the argument of the forthcoming sections. According to this point of view there is no contradiction with the above discussed Landauer principle: the gate is reversible and so we can expected that according to \virg{the experimental results presented in Fig.\til 3(a) of \cite{LNG16} the dissipated heat can be reduced below $k_B T$.} As usual in reversible logic, the value $[\alpha_i]$ set to 0 corresponds to the \emph{ancilla} input and the output values $[V(P_1)], [V(P_2)]$ are the \emph{garbage} of the logic \textsc{Or} realization by the output $[\alpha_o]$.
 %---
Situations of this kind, making use of the Toffoli's box representation introduced in \cite{To80}, can be depicted as the Figure \ref{fig:OR-CL}.

%---
%---
%\newpage
%-----------------------------------
\subsection{A formal analysis of the LNG device behavior and an interesting metatheoretical contraposition}
%-----------------------------------
In order to better understand the above (LNG) and (CL) two different points of view about the LNG experimental results, let us introduce a formalization of its behavior. First of all let us denote by $\cV$ the collection of possible voltages applied to the two probes $P_1$ and $P_2$, and by $\cA$ the collection of all possible angles assumed by the cantilever with respect to the vertical axis. From a more general point of view we can formalize the functioning of the physical device \virg{Probes+Cantilever} by a function
$$
F:\cV\times\cV\times\cA \to \cV\times\cV\times\cA
$$
assigning to the input $(V_1,V_2,\alpha_i)$ consisting of the voltage $V_1$ (resp., $V_2$) applied to the probe $P_1$ (resp., $P_2$) and the initial angle of the cantilever $\alpha_i$ the output
$$
F(V_1, V_2,\alpha_i)=(F_1(V_1,V_2,\alpha_i),F_2(V_1,V_2,\alpha_i),F_3(V_1,V_2,\alpha_i))
$$
where, as it happens in any multivalued function, the three component functions are put in clear evidence. Precisely, in order to describe the behavior of the LNG device synthesized by the previous points (Ex1)--(Ex2), the component functions $F_1:\cV\times\cV\times\cA \to\cV$, $F_2:\cV\times\cV\times\cA \to\cV$, and $F_3:\cV\times\cV\times\cA \to\cA$ are defined by the laws
$$
F_1(V_1,V_2,\alpha_i)= V_1,\;F_2(V_1,V_2,\alpha_i)=V_2,\; \text{and}\; F_3(V_1,V_2,\alpha_i)=\alpha_o
$$
In the particular case depicted in Fig.\til \ref{fig:in-out} in which the initial angle is $\alpha_i=0$ we have
\begin{align*}
&F_1(V_1,V_2,0)= V_1\\
&F_2(V_1,V_2,0)=V_2\\
&F_3(V_1,V_2,0)=\alpha_o
\end{align*}
%---
Now with respect to this formalization there are at least two possible descriptions:
\begin{enumerate}[(Po1)]
\item
According to (LNG) in considering the output results one can neglect what happens in the two probes, that is one disregards the two component functions $F_1$ and $F_2$, considered as \emph{hidden}, and so in describing the experiment one takes into account the sole function $F_3(V_1,V_2,\alpha_i)=\alpha_o$ in which the output $\alpha_0$ uniquely depends from $V_1$ and $V_2$ producing the gate of Table \ref{tbl:LNG-g} for realizing the \textsc{Or} connective in an \emph{irreversible} way.
\item
According to (CL) \emph{also} in describing the output results one must take into account the whole $\cV\times\cV\times\cA$ situation of the experimental apparatus \virg{Probes+Cantilever}, that is the whole three component  functions $F_1, F_2$, besides to $F_3$, leading to the gate described by Table \ref{tbl:CL-g} for realizing the \textsc{Or} connective in a \emph{reversible} way.
\end{enumerate}

Relatively to the above discussion we have
\begin{itemize}
\item
the theoretical Landauer principle whose validation or falsification can be obtained by experiments;
\item
the LNG experimental device which realizes the connective \textsc{Or} with an arbitrary small dissipation of energy.
\end{itemize}
So, there are two possible contradictory positions:
\begin{enumerate}[(a)]
\item
If \emph{a priori} one is against the Landauer principle, then one accepts the above position (Po1) claiming that the experimental LNG results falsify the Landauer principle.
\item
If \emph{a priori} one accepts the Landauer principle, then one agrees  with the above description (Po2) of the experimental LNG results as a corroboration of the Landauer principle.
\end{enumerate}
A very interesting situation for an epistemological/philosophical debate where the experimental results, according to one or the opposite other assumption, lead to a falsification or a corroboration of the same theoretic principle.

We of course support the position (Po2). It is out of any doubt that the experimental LNG device is formed by the physical system \virg{Probes ($P_1, P_2$)+Cantilever ($L$)} and so a formal description of its \emph{input state} must consist of  a triple $(V_1,V_2,\alpha_i)$ formed by the two input probe voltages $V_1$ and $V_2$ and the input initial angle $\alpha_i$ (left side of Figure \ref{fig:in-out}).
After the interaction the physical LNG device continue to consist of the pair \virg{Probes ($P_1, P_2$)+Cantilever ($L$)} and so in order to describe this physical situation the \emph{output state} must be formed by the complete information about not only the output angle $\alpha_o$, but also of the probe voltages (right side of Figure \ref{fig:in-out}), producing the \emph{reversible} transition described by the Table \ref{tbl:CL-g}. In other words, also in the output case we must have a \emph{complete} description of the physical state of the \virg{Probes+Cantilever} device.

On the contrary, as supported by LNG, if one decides that in the case of the output the physical system \emph{collapses} in the cantilever component disregarding what happens to the two probes, then the output state consists in the unique variable \virg{output angle} $\alpha_o$, i.e., a \emph{hidden} variables \emph{incomplete} description, corresponding to the \emph{irreversible} transition described in Table \ref{tbl:LNG-g}.

Comparing these two positions, we can say that our description is a \emph{reversible} completion of the incomplete (with hidden variables) \emph{irreversible} LNG description. This is the reason which leads us to adopt the reversible completion version of the hidden incomplete irreversible one as interesting argument of investigation in the forthcoming sections. In particular, in the next subsection we confirm the rightness of this choice just on the basis of some experimental results obtained by the LNG device.
%------------------------------------------------
\subsection{A first reversible version of the LNG device}
\label{sec:1st-LNG}
%--------------------------------------------------
 Coming back to the LNG device described at the beginning of section \ref{sec:LNG-Or} and depicted in Figg.\til \ref{fig:LNG} and \ref{fig:in-out}, owing to the fact that the input voltages $V(P_1)=0$ and $V(P_2)=0$ produce the trivial output $\alpha_o=0$, the main interesting results regard the measure of the angle deviation of the cantilever $\alpha_o$ in the three cases of input interest $[V(P_1)][V(P_2)]=01, 10, 11$. Since the cantilever is really very small, the position change of its tip, as consequence of the bent, is also very small and subjected to thermal fluctuations. Hence, the statistical distribution of the cantilever tip position is a random quantity well reproduced by a Gaussian curve. It is experimentally observed (see Fig.\til 2(b) from \cite[pag.\til 2]{LNG16}) that

\begin{itemize}
\item
logic inputs corresponding to the states $01$ e $10$ produce very \emph{similar} results, distributed in a range between 0.8 nm and 1 nm,
\item
whereas the logic input 11 produces a larger displacement around 1.1 nm.
\end{itemize}
Of course, if one agrees with these considerations then one can achieve the following LNG conclusions:
\begin{enumerate}
\item[(LNG-1)]
if the \emph{similarity} of the results obtained by the inputs 01 and 10 is assumed as an element of their \emph{indistinguishability}, considering them as the production of the same output 1, and also if the larger  displacement produced by the input 11 is always associated to the same output 1, one can conclude that \virg{the cantilever-based gate performs like an \textsc{Or} gate that is a logical irreversible device [see Table \ref{tbl:LNG-g}]: in fact there is at least one case [i.e., 01, 10, 11] where, from the sole knowledge of the logic (and physical) output [i.e., 1], it is not possible to infer the status of the logic inputs.} \cite{LNG16}.
\end{enumerate}

This is a possible metatheoretical position which can be considered as a joke of a dark night in which all the cows [inputs 01, 10, 11] result of colour black [the same output 1]. On the basis of the obtained results, our position is on the contrary quite different and in agreement with the previous considerations collected in the above statement (CL). Precisely, referring to the experimental results of Fig.\til 3(a) of \cite{LNG16}, for any fixed protocol time $\tau_p$ (ms) the average produced heat gives three different results always interpreted as the output 1 but relatively to the inputs 10 (symbol $\circ$), 01 (symbol $\vartriangle$), and 11 (symbol $\square$).

\begin{figure}[h]
	\begin{center}
		\includegraphics[width=8cm]{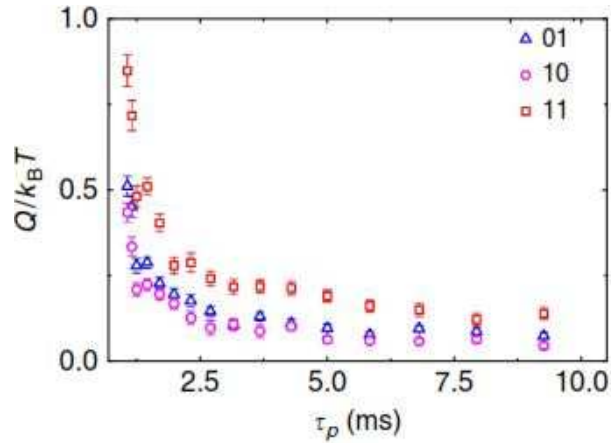}
	\end{center}
	\caption{Experimental results of Figure 3(a) from \cite[pag.\til 4]{LNG16}}
	\label{fig:3a-LNG}
\end{figure}

Once adopted the convention of identifying the following symbols
\begin{equation}
\label{eq:1-conv}
10 \equiv \circ \qquad 01\equiv\vartriangle\qquad 11\equiv \square
\end{equation}
it is really true that the experimental outputs produced by 10 and 01 are very near, as said before \virg{very similar}, between them, but as evident from the above Figure \ref{fig:3a-LNG}, reproduction of the original Figure 3(a) from \cite{LNG16},
\begin{quote}
\emph{the $\circ$ are always less in value with respect to $\vartriangle$, and furthermore the output $\square$ furnishes always a value resolutely greater that these two.}
\end{quote}

This behavior is confirmed by the histograms of Fig.\til 2(c) of \cite[pag.\til 2]{LNG16} in which the one corresponding to the input 10 ($\circ$) assumes the maximum value lightly near to the value 0.8 nm, whereas the one corresponding to the input 01 ($\vartriangle$) shows the maximum value lightly near to the value 1.0 nm, in any case greater that the previous one. Lastly, the histogram of the input 11 ($\square$) has a maximum between 1.0 and 1.2 nm, but near this latter value and in any case clearly distinguishable from the other two.

\begin{figure}[h]
	\begin{center}
		\includegraphics[width=7cm]{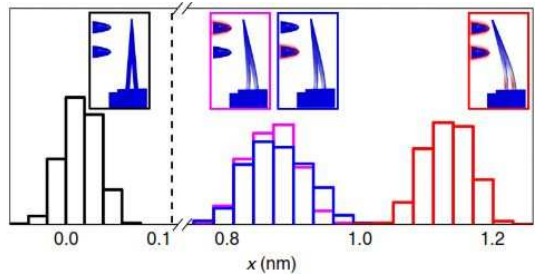}
	\end{center}
	\caption{Histograms of Figure 2(c) from \cite[pag.\til 2]{LNG16}}
	\label{fig:2c-LNG}
\end{figure}
%\newpage

In this paper we assume that they correspond to three \virg{\emph{different}} values of the logic value 1 according to the following statement:
\begin{enumerate}
\item[(CL-1)]
Borrowing the usual terminology from the fuzzy set theory according to Zadeh we can think to three different logic values 1, each of them characterized by the \virg{membership degree} represented by the three symbols  $\circ$, $\vartriangle$, and $\square$, formalized as ordered pairs $(\circ,1)$, $(\vartriangle,1)$, e $(\square,1)$.
\end{enumerate}
Then, according to this interpretation, the LNG gate for realizing the logic connective \textsc{Or} is formalized by the transitions:
\begin{equation}
\label{tabl:LNG}
\begin{tabular}{ccl}
\hline
&LNG--\textsc{Or}&
\\
\hline
Input && Output
\\
\hline
%\\
$00$ & $\Longrightarrow$ & $(*,0)$
\\
\hline
%\\
$10$ & $\Longrightarrow$ & $(\circ,1)$
\\
$01$ & $\Longrightarrow$ & $(\vartriangle,1)$
\\
$11$ & $\Longrightarrow$ & $(\square,1)$
\\
\hline
\end{tabular}
\end{equation}
where for formal completeness we used the symbol $*$ associated to the logic value 0 in order to obtain the output $(*,0)$, omitting its precise determination which will be made in the sequel. In this way we obtain a \emph{reversible} gate since the knowledge of any output allows one to uniquely determine the corresponding input generating it. In this way we lose any possible ambiguity, and with respect to this result we can do the following remark.
  %and so it turns out quite natural the satisfaction of the Landauer principle of an arbitrary small energy dissipation.
\begin{quote}
\emph{It is very interesting to note that these pure experimental results are completely described in a right way by (i.e., it is totally coincident with) our Table \ref{tbl:CL-g} once adopted in this latter the above conventional substitutions given by equation \eqref{eq:1-conv}. In this way the experimental results of Fig.\til 2(b) from \cite{LNG16}) confirm the rightness of our previous assumptions formalized in (Po2) relative to a 3in/3out reversible gate description of the LNG experiment, and so without any contradiction with respect to the experimental result of arbitrary small dispersion of energy, according to the Landauer principle.}
\end{quote}

At any rate we don't continue to develop this interesting analysis since in the next section we formalize a realization of the LNG device as a 3in/3out self--reversible gate, avoiding any discussion about the dichotomy \virg{01 and 10 distinguishable or indistinguishable}, but which satisfies the Landauer principle owing to its reversibility.
%--------------------------------
\section{The self--reversible 3in/3out Cattaneo Leporini (CL) gate}
\label{sec:3-3CL-LNG}
%---------------------------
In order to obtain this result first of all we have analyzed the main 3in/3out gates which one can found in literature: the conservative self--reversible Fredkin gate, the self--reversible but not conservative Toffoli and Peres gates (see \cite{FT82, Pe85}) realizing that no of them satisfies the condition of having as \emph{derived gate}, that is fixing one of the input as ancilla and considering two of the outputs as garbage, the description of the LNG device. As a consequence we have autonomously construct a gate of this kind arriving to the Cattaneo--Leporini (CL) 3in/3out gate  $G_{\text{CL}}$  described by the following functional representation (but discovering some times after our formalization that the same gate has been introduced in \cite{KTR12} with the name of \textsc{TNor} gate).
\begin{align}
\label{eq:CLV-fun}
\text{CL Gate} =
\begin{cases}
x'_1 = x_1 \\
x'_2 = x_2 \\
x'_3 = (x_1\lor x_2)\oplus x_3
\end{cases}
\end{align}

This functional definition is represented by the Table \ref{tbl:CLV-gate}.

%\begin{equation}
\begin{table}[h]
      \begin{tabular}{|c|c||c||c||c|c||c||}
         \hline
         $x_1$ & $x_2$ & $x_3$ & $\xrightarrow{G_{CL}}$ &
                 $x'_1$ & $x'_2$ & $x'_3$ \\
         \hline\hline
         \mbox{\rule[0cm]{0cm}{2.5ex}
         $0$}& $0$ & $0$ & & $0$ & $0$ & $0$\\
         $0$ & $0$ & $1$ & & $0$ & $0$ & $1$ \\
         $0$ & $1$ & $0$ & & $0$ & $1$ & $1$ \\
         $0$ & $1$ & $1$ & & $0$ & $1$ & $0$ \\
         \hline
         $1$ & $0$ & $0$ & & $1$ & $0$ & $1$ \\
         $1$ & $0$ & $1$ & & $1$ & $0$ & $0$ \\
         $1$ & $1$ & $0$ & & $1$ & $1$ & $1$ \\
         \mbox{\rule[0cm]{0cm}{1.5ex}$1$}
             & $1$ & $1$ & & $1$ & $1$ & $0$\\
         \hline
      \end{tabular}
      \vspace{0.3cm}
      \caption{Tabular representation of the 3in/3out self-reversible CL gate}
\label{tbl:CLV-gate}
\end{table}
% \end{equation}
%\newpage

Obviously, this is a reversible non--conservative gate (for instance, in the transition  $(010)\to (011)$ the number of 1 bites in the input is not preserved). Moreover, it is self--reversible in the sense that
\begin{equation}
\label{eq:CLV-self-rever}
G_{\text{CL}}\circ G_{\text{CL}} = \text{id}
\end{equation}
%----
That is, the inverse of  $G_{\text{CL}}$ is the $G_{\text{CL}}$  itself:  $G_{\text{CL}}^{-1}=G_{\text{CL}}$. Below in Fig.\til \ref{fig:bi-CLV} it is represented the role played by self--reversibility in producing as global output the same input as consequence of the transitions  $(a,b,c)\xrightarrow{CL} (a,b,(a\lor b)\oplus c) \xrightarrow{CL} (a,b,c)$. The \textsc{Fan}\textsc{Out} (FO) reversible, but non conservative, gate allows the duplication (cloning) of the signal of the third line after the first output, a duplication of which is inserted as third input of the second CL gate, whereas the other duplication is extracted as overall output of the cascade.

\begin{figure}[ht]
\begin{center}
   \includegraphics[width=10cm]{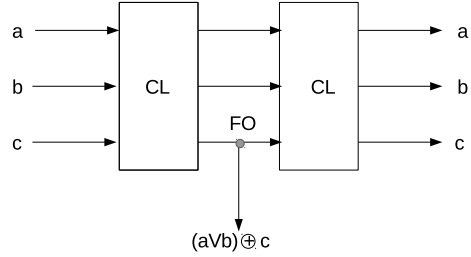}
\end{center}
   \vspace{-2.0cm}
      \caption{Self--reversibility of the CL gate, with the FO gate for the cloning of the third line of the first output}
      \label{fig:bi-CLV}
\end{figure}

From the functioning of the CL gate described in the Table \ref{tbl:CLV-gate} it follows that, if as usual some suitable inputs are fixed as ancilla, one has the following possible cases:

 \begin{align*}
 (0,a,b)&\to (0,a,a\oplus b) & (a,0,b)&\to (a,0,a\oplus b)   & (a,b,0)&\to (a, b,a\lor b)\\
 (1,a,b)&\to (1,a,\neg b)    & (a,1,b)&\to (a,1,\neg b)      & (a,b,1)&\to (a, b,\neg(a\lor b))
 \end{align*}
%---
corresponding to the realization of the logic connectives \textsc{Xor} $\oplus$, \textsc{Or} $\lor$, and \textsc{Nor} $\neg\lor$. Moreover, below one can find some realizations of the  \textsc{Fan}\textsc{Out} gate:
$$
(a,0,0)\to (a,0,a)\qquad (0,a,0)\to (0,a,a)
$$
%---
There are also some different possibilities of realizing the logic connective \textsc{Not} $\neg$, of which below we show some of them:
%---
\begin{subequations}
\label{eq:Not-CL}
\begin{gather}
(0,1,b)\to (0,1,\neg b)\qquad (1,0,b)\to (1,0,\neg b)\qquad (1,1,b)\to (1,1,\neg b)
\label{eq:Not-CL-a}
\\
(a,0,1)\to (a,0,\neg a)\qquad    (0,b,1)\to (0,b,\neg b)
\end{gather}
\end{subequations}
%---
Summarizing, the primitive logic connectives which can be obtained from this 3in/3out self--reversible CL gate can be collected in the following list:
%---
$$
\mathcal{G}_C=\{\textsc{Xor},\textsc{Or},\textsc{Nor},\textsc{Not}, \textsc{Fan}\textsc{Out}\}
$$

Let us note a behavior of this CL gate which in some sense is dual with respect to the Toffoli gate, as successively discussed by a comparison of it with this latter, and of which we will study the analogies in the forthcoming section. This behavior consists in the transitions:
%---
\begin{equation}
\label{eq:noi-CCnot}
(0,0,c)\xrightarrow{a\land b =0} (0,0,c)\qquad (a,b,c)\xrightarrow{a\lor b =1} (a,b,\neg c)
\end{equation}
%---
in which we stress that if the first and the second lines are both fixed in the input 0 ($a\land b=0$) then on the third line it acts the identity, whereas in all the other cases in which at least one of the two control lines, the first and the second ones, is fixed with the input 1 ($a\lor b=1$) then on the third line it is the \textsc{Not} gate which acts. It is a kind of Controlled--Controlled--(multiple)Not (CCmN) in which $a$ and $b$ are known as the first and second \emph{control bits}, while $c$ is the \emph{target bit}. In conclusion, the gate leaves both control bits unchanged, flips the target bit if at least one of the control bit is set to 1, and otherwise leaves the target bit alone. We speak of \emph{multipleNot} since we have seen in the second transition of the equation \eqref{eq:noi-CCnot} how it is possible to generate in three different modes the \textsc{Not} logic gate when at least one of the inputs $a$ or $b$ is fixed in the bit 1 (see also the equations \eqref{eq:Not-CL-a}).

Let us now analyze the behavior of our self--reversible CL gate when the third input is fixed to $x_3=0$, which turns out to be useful in the sequel for a comparison with the LNG realization of their \textsc{Or} gate. We extract this behavior from the Table \ref{tbl:CLV-gate} when the third input is set to 0, obtaining the partial Table \ref{tab:CL-c=0} (which is identical to the Table \ref{tbl:CL-g} under the identifications $x_1=x'_1=[V(P_1)]$, $x_2=x'_2 = [V(P_2)]$, and $x_3=[\alpha_i]$, $x'_3= [\alpha_o]$):

%---
%\begin{equation}
\begin{table}[h]
      \begin{tabular}{|c|c||c||c|c|c||c||}
         \hline
         $x_1$ & $x_2$ & $x_3$ & $\xrightarrow{G_{CL}}$ &
                 $x'_1$ & $x'_2$ & $x'_3$ \\
         \hline\hline
         \mbox{\rule[0cm]{0cm}{2.5ex}
         $0$}& $0$ & $0$ & & $0$ & $0$ & $0$\\
         $0$ & $1$ & $0$ & & $0$ & $1$ & $1$ \\
         $1$ & $0$ & $0$ & & $1$ & $0$ & $1$ \\
         \mbox{\rule[0cm]{0cm}{1.5ex}$1$}
             & $1$ & $0$ & & $1$ & $1$ & $1$ \\
         \hline
      \end{tabular}
%---
\vspace{0.3cm}
\caption{Realization of the \textsc{Or} connective by the output $x'_3=x_1 \lor x_2$ from the CL self-reversible 3in/3out gate fixing the input $x_3=0$}
\label{tab:CL-c=0}
%---
\end{table}
\newpage
%\end{equation}

\begin{figure}[ht]
\begin{center}
   \includegraphics[width=8cm]{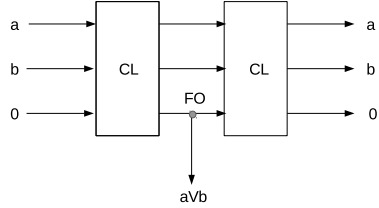}
\end{center}
   \vspace{-1.5cm}
      \caption{Self--reversibility of the CL gate for $x_3=0$, with the FO gate for the cloning of the third output of the first gate generating the \textsc{Or} logic connective}
      \label{fig:bi-CLV-OR}
\end{figure}

This situation produces the self--reversible transitions:
 %---
 $$
 (a,b,0)\to (a,b,a\lor b)\to (a,b,0)
 $$
represented in the block scheme of Figure \ref{fig:bi-CLV-OR}.

%---
%==================================
\begin{comment}
Thus, using twice the CL gate with the input $x_3=0$, we have the explicit self-reversible transitions:
\begin{equation*}
   %\label{tbl:CLV-gate-30}
%---
      \begin{tabular}{ccccc}
         \hline
      {} &CL&{} &CL&{} \\
         \hline
      000&$\to$&000&$\to$&000\\
      010&$\to$&011&$\to$&010\\
      100&$\to$&101&$\to$&100\\
         \mbox{\rule[0cm]{0cm}{1.5ex}110}
             & $\to$ & 111 & $\to$ & 110 \\
         \hline
\end{tabular}
\end{equation*}
\end{comment}
%========================================
%\newpage

\begin{figure}[ht]
\begin{center}
   \includegraphics[width=8cm]{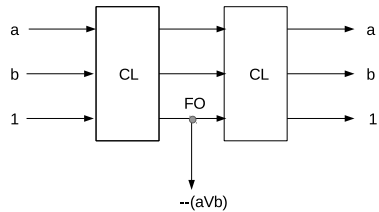}
\end{center}
   \vspace{-1.5cm}
      \caption{Self--reversibility of the CL gate for $x_3=1$, with the FO gate for the cloning of the third output of the first gate generating the \textsc{Nor} logic connective}
      \label{fig:bi-CLV-NOR}
\end{figure}

Note that if on the contrary in the CL gate of Table \ref{tbl:CLV-gate} the third input is fixed to 1, $x_3=1$, then both the first and the second line produce the identity whereas the third output furnishes the self-reversible \textsc{Nor} gate realization according to the transitions:
$$
(a,b,1) \to (a,b,\neg(a\lor b))\to (a,b,1)
$$
depicted in the Figure \ref{fig:bi-CLV-NOR}.
%---
%\newpage

Let us finish this section with the analysis of another possible behavior of the CL gate. Precisely the one in which
   %---This CL gate turns out to be a \virg{controlled gate} in the sense of a gate in which
the first line has an output which remains unchanged  $x'_1 = x_1$, but with the two input bits 0 and 1 which produce two different situations acting as different gates on the remaining two lines $x_2 - x'_2$ e $x_3- x'_3$ according to the following tables.

\begin{equation}
\label{tbl:CL-contr}
      \begin{tabular}{|l||c|c|c|}
         \hline
      fixed line & 0 &$\text{id}$ & 0\\
      \hline
      first line   & a &$\text{id}$ & a\\
      second line  & b &$\oplus$    & $a\oplus b$\\
      \hline
     \end{tabular}
\qquad
%---
      \begin{tabular}{|l||c|c|c|}
         \hline
      fixed line & 1 &$\text{id}$ & 1\\
      \hline
      first line   & a &$\text{id}$ & a\\
      second line  & b &$\neg$   & $\neg b$\\
      \hline
     \end{tabular}
\end{equation}
%--------------------------------
\section{The self--reversible 3in/3out Toffoli (T) gate}
\label{sec:3-3T}
%--------------------
In literature one can find an interesting reversible, non conservative, gate introduced by Toffoli by the functional representation:
\begin{equation}
\label{eq:T-fun}
\text{Toffoli Gate} =
\begin{cases}
x'_1 = x_1\\
x'_2 = x_2 \\
x'_3 = (x_1\land x_2)\oplus  x_3
\end{cases}
\end{equation}
%---
whose comparison with the functional representation of the CL gate of equation \eqref{eq:CLV-fun} stressed the difference consisting in substituting into the third equation of this latter the connective $\lor$ with the connective $\land$. The Toffoli gate formulation in terms of truth table is the following one:
%---

%\begin{equation}
\begin{table}[h]
      \begin{tabular}{|c|c||c||c||c|c||c||}
         \hline
         $x_1$ & $x_2$ & $x_3$ & $\xrightarrow{G_T}$ &
                 $x'_1$ & $x'_2$ & $x'_3$ \\
         \hline\hline
         \mbox{\rule[0cm]{0cm}{2.5ex}
         $0$}& $0$ & $0$ & & $0$ & $0$ & $0$\\
         $0$ & $0$ & $1$ & & $0$ & $0$ & $1$ \\
         $0$ & $1$ & $0$ & & $0$ & $1$ & $0$ \\
         $0$ & $1$ & $1$ & & $0$ & $1$ & $1$ \\
         \hline
         $1$ & $0$ & $0$ & & $1$ & $0$ & $0$ \\
         $1$ & $0$ & $1$ & & $1$ & $0$ & $1$ \\
         $1$ & $1$ & $0$ & & $1$ & $1$ & $1$ \\
         \mbox{\rule[0cm]{0cm}{1.5ex}$1$}
             & $1$ & $1$ & & $1$ & $1$ & $0$\\
         \hline
      \end{tabular}
      \vspace{0.3cm}
      \caption{Tabular representation of the 3in/3out self-reversible Toffoli gate}
\label{tbl:T-gate}
\end{table}
%\end{equation}
%\newpage

 From which, by constraining one of the inputs as ancilla, it is possible to obtain some familiar standard logic primitives, according to the correspondences
 \begin{align*}
 (0,a,b)&\to (0,a, b)          & (a,0,b)&\to (a,0,b)                & (a,b,0)&\to (a,b,a\land b)\\
 (1,a,b)&\to (1,a,(a\oplus b)) & (a,1,b)&\to (a, 1, a\oplus b) & (a,b,1)&\to (a, b, \neg(a\land b))
 \end{align*}
 whereas, by constraining two of the inputs, we may get the \textsc{Fan}\textsc{Out} and the \textsc{Not} gates, for instance
 $$
 (a,1,0)\to (a,1,a)\qquad (1,b,0)\to (1,b,b)\qquad (a,1,1)\to (a,1,\neg a)\qquad (1,b,1)\to (1,b,\neg b)
 $$
Summarizing the set of logic primitives generated by the self--reversible Toffoli gate is the following one:
 $$
 \mathcal{G}_T=\parg{\textsc{Xor},\textsc{And},\textsc{Nand},\textsc{Not},\textsc{Fan}\textsc{Out}}
 $$

  Let us recall that this Toffoli gate is the one that Feynman in \cite{Fe96} defines (and presently is recognized as such) \textsc{Controlled} \textsc{Controlled} \textsc{Not} (CCN): \virg{in which the lines $x_1$ and $x_2$ acts as control lines, leaving $x_3$  as it is unless both are one, in which case $x_3$ becomes \textsc{Not}($x_3$).}
This behavior can be described by the correspondences, which can be compared with the analogous of the CL gate expressed by equation \eqref{eq:noi-CCnot},
%---
\begin{equation}
\label{eq:T-CCnot}
(a,b,c)\xrightarrow{a\lor b =0} (a,b,c)\qquad (1,1,c)\xrightarrow{a\land b =1} (1,1,\neg c)
\end{equation}

Similarly to the CL gate, fixing the input $x_3=0$ as ancilla, the Toffoli gate reduces to the following table producing the output $x'_3=x_1\land x_2$ with the pair $x'_1, x'_2$ as garbage, i.e., a reversible realization of the \textsc{And} logic connective.
%---
%\begin{equation}
\begin{table}[h]
      \begin{tabular}{|c|c||c||c|c|c||c||}
         \hline
         $x_1$ & $x_2$ & $x_3$ & $\xrightarrow{G_T}$ &
                 $x'_1$ & $x'_2$ & $x'_3$ \\
         \hline\hline
         \mbox{\rule[0cm]{0cm}{2.5ex}
         $0$}& $0$ & $0$ & & $0$ & $0$ & $0$\\
         $0$ & $1$ & $0$ & & $0$ & $1$ & $0$ \\
         $1$ & $0$ & $0$ & & $1$ & $0$ & $0$ \\
         \mbox{\rule[0cm]{0cm}{1.5ex}$1$}
             & $1$ & $0$ & & $1$ & $1$ & $1$ \\
         \hline
      \end{tabular}
      \vspace{0.3cm}
\caption{Realization of the \textsc{And} connective by the output $x'_3=x_1 \land x_2$ from the Toffoli self-reversible 3in/3out gate fixing the input $x_3=0$}
\label{tab:T-c=0}
\end{table}
%---
%\end{equation}

This situation of the Toffoli gate produces the self-reversible transitions:
$$
(a,b,0)\to (a,b,a\land b)\to (a,b,0)
$$

 The logic connective \textsc{Nand}, $\neg(a\land b)$, is obtained by the Toffoli gate if we input $x_1=a$ and $x_2=b$ as control bits, and fixing as ancilla bit the third input  $x_3=1$. The \textsc{Nand} of $x_1$ and $x_2$ is output as the target $x'_3$ considering the pair formed by the first and the second outputs  $(x'_1=a, x'_2=b)$ as garbage. Formally, this can be formalized by the transitions, where the second one stresses the self-reversibility of the gate:
 $$
 (a,b,1)\to (a,b,\neg(a\land b))\to (a,b,1)
 $$
 Quoting Peres \virg{It is well known that the \textsc{Nand} gate is a universal primitive (\textsc{Nand} is the relational operator which gives the "true'' value as output if one or both input values are "false.'') Therefore the reversible gate [of Table \ref{tbl:T-gate}] is universal.}
%--------------------
\section{LNG \textsc{Or} and \textsc{And} connectives implementation by self-reversible gates}
%\section{Realizzazione della porta \textsc{Or} tramite LNG e realizzazione secondo CLV}
\label{sec:Or-LNG-CLV}
%--------------------
Let us first consider the \textsc{Or} gate realization using the device proposed by L\'opez-Su\'arez et al.\til in \cite{LNG16} (schematically depicted in Fig.\til \ref{fig:LNG} of our section \ref{sec:LNG-Or}).
As previously described the device consists of two electrical probes (denoted by $P_1$, $P_2$) close to a cantilever $L$.
%---
An electrode voltage $V$ can be applied to each probe $P_j$.
If $V=0$, we will say the probe is off (denoted by $D$), while we will say the probe is on (denoted by $A$) when $V\neq 0$.
The corresponding cantilever tip displacement is determined by the angle $\alpha$ formed with respect to the vertical axis as reference axis.

The experimental results obtained by the LNC device represented by Fig.\til 2(c) from \cite{LNG16} can be summarized in the following points:
\begin{enumerate}
\item[(Exp1)]
Under the fixed input initial angle $\alpha_i=0$ of the cantilever, a sufficiently large number of single tests, each of which relative to the three non \virg{trivial} inputs of the voltages applied to the two probes $DA$, $AD$, and $AA$, experimentally produces three histograms $I_{DA}$, $I_{AD}$, and $I_{AA}$ as mappings $\alpha\in\IR_+ \to I_{hk}(\alpha)\in\IN$, for $hk$ ranging on $\{DA, AD, AA\}$.
\item[(Exp2)]
Each histogram has a support (collection of $\alpha\in\IR_+$ for which the corresponding $I_{hk}(\alpha)\neq 0$), denoted as $supp(I_{hk})$, which is a bounded interval on $\IR_+$.
\item[(Exp3)]
There is a bounded value of angle $\alpha_B$, such that both the supports $I_{DA}<\alpha_B$ and $I_{AD}<\alpha_B$, whereas $\alpha_B< I_{AA}$.
\item[(Exp4)]
The two supports $I_{DA}$ and $I_{AD}$ are quite \emph{similar}, $I_{DA}\simeq I_{AD}$, and so according to the LNG assumption described in the point (LNG-1) of section \ref{sec:1st-LNG}, they can be submitted to the stronger condition of being equal:  $I_{DA}= I_{AD}$.
\item[(Exp5)]
All the above histograms $I_{hk}$ correspond to  statistical distributions well reproduced by Gaussian curve. Denoted by $\hat\alpha_1 = <I_{DA}>$, $\tilde{\alpha}_1=<I_{AD}>$, and $\alpha_2=<I_{AA}>$, the corresponding mean value of the involved Gaussian, ad as usual in physics putting the boundary angle $\alpha_B=1$, the experimental results give the following chain of values:
\\
$
0< \hat{\alpha}_1 \simeq \tilde{\alpha}_1 < 1 < \alpha_2
$.
\item[(Exp6)]
According to the strong assumption of point (Exp4), this last result can be formalized by the chain of inequalities:
\begin{equation}
\label{eq:ha1=ta1}
0< \hat{\alpha}_1 = \tilde{\alpha}_1 < 1 < \alpha_2
\end{equation}
\end{enumerate}

We recall that all these experimental results follow by the initial condition on the input angle $\alpha_i=0$, and so all these results can be collected in the following table where, according to the strong assumption (Exp6) formalized in the chain of inequalities \eqref{eq:ha1=ta1}, we set $\alpha_1:=\hat{\alpha}_1=\tilde{\alpha}_1$.

%---\begin{equation}
\begin{table}[h]
%---
      \begin{tabular}{|c|c||c||c|c|c||c||}
         \hline
         $P_1$ & $P_2$ & $\alpha_i$ & $\longmapsto$ &
                 $P'_1$ & $P'_2$ & $\alpha_o$ \\
         \hline\hline
         \mbox{\rule[0cm]{0cm}{2.5ex}
         $D$}& $D$ & $0$ & & $D$ & $D$ & $0$\\
         $D$ & $A$ & $0$ & & $D$ & $A$ & $\alpha_1$ \\
         $A$ & $D$ & $0$ & & $A$ & $D$ & $\alpha_1$ \\
         \mbox{\rule[0cm]{0cm}{1.5ex}$A$}
             & $A$ & $0$ & & $A$ & $A$ & $\alpha_2$ \\
         \hline
      \end{tabular}
      \vspace{0.3cm}
      \caption{Physical behavior of the LNG device when the cantilever is in the initial vertical position $\alpha_i=0$ and the ordering of the final angles $\alpha_o$ given by $0<\alpha_1< 1 <\alpha_2$}
\label{tbl:Ex-LNG-a0}
\end{table}
%---\end{equation}

Assuming now the convention of putting $A=1$ (probe on, i.e., active) and $D=0$ (probe off, i.e., inactive) these results can be translated in a Boolean context under  particular assumptions about a \emph{normalization function} of the  angle $\alpha$ according to two possible cases which we will discuss now.
%-------------------
\subsection{First normalization function: the Cattaneo-Leporini case for generating \textsc{Or}}
\label{ssec:CL-norm-Or}
%-------------------
In this first case the normalization function of the angle, denoted by $u_1(\alpha)$, is the following one:
$$
u_1(\alpha)= \begin{cases}
0 & \text{if}\; \alpha=0\\
1 & \text{otherwise}
            \end{cases}
$$

From this choice of normalization function, the physical behavior of the LNG device described in the Table \ref{tbl:Ex-LNG-a0} when the initial cantilever position is vertical, $\alpha_i=0$, is translated into the table of equation \eqref{tbl:CLV-gate-30} at the left side.
The Boolean table at the right is nothing else than this latter setting $A=1$, $D=0$, $P_1=x_1$, $P_2=x_2$, $u_1(\alpha_i)=x_3$, and $u_1(\alpha_o)=x'_3$:

\begin{equation}
%---\begin{table}
\label{tbl:CLV-gate-30}
%---
      \begin{tabular}{|c|c||c||c|c|c||c||}
         \hline
         $P_1$ & $P_2$ & $u_1(\alpha_i)$ & $\longmapsto$ &
                 $P'_1$ & $P'_2$ & $u_1(\alpha_o)$ \\
         \hline\hline
         \mbox{\rule[0cm]{0cm}{2.5ex}
         $D$}& $D$ & $0$ & & $D$ & $D$ & $0$\\
         %\hline
         $D$ & $A$ & $0$ & & $D$ & $A$ & $1$ \\
         %---$0$ & $1$ & $1$ & & $0$ & $1$ & $0$ \\
         %---\hline
         $A$ & $D$ & $0$ & & $A$ & $D$ & $1$ \\
         %---$1$ & $0$ & $1$ & & $1$ & $0$ & $0$ \\
         \mbox{\rule[0cm]{0cm}{1.5ex}$A$}
             & $A$ & $0$ & & $A$ & $A$ & $1$ \\
         %---    & $1$ & $1$ & & $1$ & $1$ & $0$\\
         \hline
      \end{tabular}
      %\caption{Normalized physical behavior of the LNG device with cantilever in the vertical initial configuration}
%---\end{table}
%---
\qquad
%---
%---\begin{table}
      \begin{tabular}{|c|c||c||c|c|c||c||}
         \hline
         $x_1$ & $x_2$ & $x_3$ & $\longmapsto$ &
                 $x'_1$ & $x'_2$ & $x'_3$ \\
         \hline\hline
         \mbox{\rule[0cm]{0cm}{2.5ex}}
         $0$ & $0$ & $0$ & & $0$ & $0$ & $0$\\
         %---$0$ & $0$ & $1$ & & $0$ & $0$ & $1$ \\
         $0$ & $1$ & $0$ & & $0$ & $1$ & $1$ \\
         %---$0$ & $1$ & $1$ & & $0$ & $1$ & $0$ \\
         %---\hline
         $1$ & $0$ & $0$ & & $1$ & $0$ & $1$ \\
         %---$1$ & $0$ & $1$ & & $1$ & $0$ & $0$ \\
         \mbox{\rule[0cm]{0cm}{1.5ex}$1$}
             & $1$ & $0$ & & $1$ & $1$ & $1$ \\
         %---    & $1$ & $1$ & & $1$ & $1$ & $0$\\
         \hline
      \end{tabular}
      %\caption{Boolean version of the LNG device presented at the left}
%---\end{table}
%---
\end{equation}

But, looking at these experimental results from the LNG device we con set the following interesting considerations:
\emph{
\begin{itemize}
\item
The table at the right giving the Boolean formalization of the LNG device behavior when the initial cantilever angle is 0 coincides with the Table \ref{tab:CL-c=0} discussed in section \ref{sec:3-3CL-LNG} as self-reversible CL gate realization of the classical \textsc{Or} connective under the ancilla choice $x_3=0$, the third output $x'_3 = x_1\lor x_2$ producing the required \textsc{Or}, and with the two outputs $x'_1$ and $x'_2$ as garbage (see also the description given by Fig.\til \ref{fig:OR-CL} and \ref{fig:bi-CLV-OR}).
\item
That is, all the experimental results obtained by the LNG concrete device and collected in the above Table \ref{tbl:Ex-LNG-a0} can be described by a 3in/3out self-reversible Boolean gate with a complete agreement with the Landauer principle of arbitrary small energy dissipation.
\item
In other words, there is no experimental contradiction as claimed by LNG in their paper.
\end{itemize}
}
%---

Furthermore, encoding the output $(x'_1, x'_2)$ as follows
$
00 = *,\, 01 =\vartriangle,\, 10=\circ,\, 11=\square
$,
the table \eqref{tbl:CLV-gate-30} at the right side becomes the following (that can be compared with Table \eqref{tabl:LNG}, according to our point of view about the experimental results obtained from the LNG device described in section \ref{sec:1st-LNG}):
%---
\begin{equation}
%---
      \begin{tabular}{|c||c||c|c||c||}
         \hline
         $x_1x_2$ & $x_3$ & $\longmapsto$ &
                 $x'_1x'_2$ & $x'_3$ \\
         \hline\hline
         \mbox{\rule[0cm]{0cm}{2.5ex}
         $00$}& $0$ & & $*$ & $0$\\
         \hline
         $01$ & $0$ & & $\vartriangle$ & $1$ \\
         $10$ & $0$ & & $\circ$ & $1$ \\
         \mbox{\rule[0cm]{0cm}{1.5ex}$11$}
             & $0$ & & $\square$ & $1$ \\
         \hline
      \end{tabular}
%---
 \end{equation}
Therefore, if in the left table \eqref{tbl:CLV-gate-30} we consider the case where at least one of the probes is active, we will have the following three transitions:
% Pertanto, se nella tabella "fisica'' a sinistra della \eqref{tbl:CLV-gate-30} consideriamo il caso in cui almeno una della punte sia attiva, avremo le seguenti tre transizioni "fisiche''
 \begin{align*}
 &(DA,u_1(\alpha_i)=0) \xrightarrow{LNG} (DA,u_1(\alpha_o)=1)\\
 &(AD,u_1(\alpha_i)=0) \xrightarrow{LNG} (AD,u_1(\alpha_o)=1)\\
 &(AA,u_1(\alpha_i)=0) \xrightarrow{LNG} (AA,u_1(\alpha_o)=1)
 \end{align*}
% in cui,
In these cases, when the result of the interaction of the probes $P_1, P_2$ on the cantilever $L$ is observed, one cannot overlook/disregard the two probes to only look at the tip position, but one has to consider the whole apparatus \virg{probes + cantilever tip} (see Fig.\til \ref{fig:in-out}).
%============================
\begin{comment}
\begin{figure}[h]
\begin{center}
   \includegraphics[width=8cm]{in-out.eps}
\end{center}
   \vspace{-1.0cm}
      \caption{Schematic description of the experimental LNG-device where at least one of the probes ($P_1$, $P_2$) is active. Both probes do not change their state and the cantilever $L$ undergoes a deflection of an $\alpha\neq 0$.}
      \label{fig:in-out}
\end{figure}
%\newpage
\end{comment}
%=============================

Following the right table \eqref{tbl:CLV-gate-30}, we have the whole transitions:
%---
 \begin{align*}
 &(01|0)\xrightarrow[CL]{LNG}(01|1)=(\vartriangle|1)\\
 &(10|0)\xrightarrow[CL]{LNG}(10|1)=(\circ|1)\\
 &(11|0)\xrightarrow[CL]{LNG}(11|1)=(\square|1)
 \end{align*}
 %---
where the output in $\{0,1\}^3$ (whether it is $(01|1)$ or $(10|1)$ or $(11|1)$) uniquely determines the input in $\{0,1\}^3$ generating it, and so, stressing this conclusion another time, according to the Landauer principle without any dissipation of energy.
All this has nothing to do with the 2in/1out irreversible \textsc{Or} logic gate where the output $1$ does not allow one to determine the generating input in $\{0,1\}^2$, as claimed by LNG in their paper.
%\newpage

As a summary of all the above discussion we can state the

\begin{description}
\item[1st Conclusion]\emph{
The LNG device under the assumptions of the initial cantilever angle $\alpha_i=0$ and the normalized function on angles $u_1$ is a concrete realization of the CL 3in/3out self-reversible gate with the third input fixed to 0, producing in the third output the connective \textsc{Or}.}
\end{description}

%------------------
\subsection{Second normalization function: the Toffoli case for generating \textsc{And}}
\label{ssec:T-norm-And}
%------------------
As said before, the Boolean formulation of the experimental results collected in Table \ref{tbl:Ex-LNG-a0} depends from an arbitrary choice of a normalization function assigning in a conventional way Boolean values to the cantilever angles. In the present subsection we take into account another possible conventional choice formalized by the normalization function:
$$
u_2(\alpha)= \begin{cases}
0 & \text{if}\; 0\le\alpha \le 1\\
1 & \text{otherwise}
            \end{cases}
$$
In this case the Table \ref{tbl:Ex-LNG-a0}, with the usual adopted conventions, leads to the following two tables
\begin{equation}
\label{tbl:CL-gate-T}
%---
      \begin{tabular}{|c|c||c||c|c|c||c||}
         \hline
         $P_1$ & $P_2$ & $u_2(\alpha_i)$ & $\longmapsto$ &
                 $P'_1$ & $P'_2$ & $u_2(\alpha_o)$ \\
         \hline\hline
         \mbox{\rule[0cm]{0cm}{2.5ex}
         $D$}& $D$ & $0$ & & $D$ & $D$ & $0$\\
         $D$ & $A$ & $0$ & & $D$ & $A$ & $0$ \\
         $A$ & $D$ & $0$ & & $A$ & $D$ & $0$ \\
         \mbox{\rule[0cm]{0cm}{1.5ex}$A$}
             & $A$ & $0$ & & $A$ & $A$ & $1$ \\
         \hline
      %\caption{Andamento fisico del dispositivo LNG quando la lamella \e inizialmente in posizione verticale}
      \end{tabular}
%---
\qquad
%---
      \begin{tabular}{|c|c||c||c|c|c||c||}
         \hline
         $x_1$ & $x_2$ & $x_3$ & $\longmapsto$ &
                 $x'_1$ & $x'_2$ & $x'_3$ \\
         \hline\hline
         \mbox{\rule[0cm]{0cm}{2.5ex}}
         $0$ & $0$ & $0$ & & $0$ & $0$ & $0$\\
         $0$ & $1$ & $0$ & & $0$ & $1$ & $0$ \\
         $1$ & $0$ & $0$ & & $1$ & $0$ & $0$ \\
         \mbox{\rule[0cm]{0cm}{1.5ex}$1$}
             & $1$ & $0$ & & $1$ & $1$ & $1$ \\
         \hline
      \end{tabular}
%---
 \end{equation}

 The table at the right coincides with the Table \ref{tbl:T-gate} obtained from the self-reversible 3in/3out Toffoli gate fixing the input ancilla $x_3=0$, with the pair $x'_1 x'_2$ considered as garbage, and the output $x'_3 = x_1\land x_2$ furnishing the logic \textsc{And} of the first and second inputs. So also in this case we have a \virg{reversible} generation of the required \textsc{And} gate with arbitrary small dissipation of energy.
 This result involving the self-reversible Toffoli gate has been achieved adopting a normalization function for describing the experimental results produced by the LNG device different from the one adopted in the CL case. But let us stress that these two different choices correspond to some quite arbitrary Boolean assignments to the involved cantilever angles, no one of which can be considered as a privileged choice from the experimental point of view.

 This behavior has been also realized by LNC when in the discussion about the Fig.\til 2(c) they assert that \virg{the threshold value for the \textsc{Or} gate is represented by the dashed line [around 0.1 nm]. By changing the position of the dashed line [around 1.0 nm], the gate can be operated also as an \textsc{And} gate.} Precisely, the dashed line around 0.1 nm put the three situations 01, 10, and 11 as associated to the output state 1, whereas the dashed line around 1.0 nm groups the three situations 00, 01, and 10 as associated to the output state 0.

As a summary of the discussion performed in this subsection we can state the

\begin{description}
\item[2nd Conclusion]\emph{
The LNG device under the assumptions of the initial cantilever angle $\alpha_i=0$ and the normalized function on angles $u_2$ is a concrete realization of the Toffoli 3in/3out self-reversible gate with the third input fixed to 0, producing in the third output the connective \textsc{And}.}
\end{description}
%---------------
\section{\textsc{Nor} and \textsc{Nand} connectives implementations by LNG device}

Let us recall that, as stressed before, the assignment of a Boolean bit value, either 0 or 1, to a deflection angle formed by the cantilever is only a conventional matter of fact. There is no physical reason to state that, for instance, a particular angle $\alpha$ may be labelled with the bit value 1 instead of the bit value 0. This is the reason that allowed us to consider two different normalization functions, $u_1$ and $u_2$, to treat the above cases of subsections \ref{ssec:CL-norm-Or} and \ref{ssec:T-norm-And}, respectively, in order to prove that the LNG device, suitably normalized, describes the \textsc{Or} and the \textsc{And} logic connectives by self-reversible CL and Toffoli gates.
%-----------------
\subsection{Third normalization function: the Cattaneo-Leporini case for generating \textsc{Nor}}
%---------------
Let us apply to the functioning of the LNG device described by the Table \ref{tbl:Ex-LNG-a0} the conventional normalization function $\Ov u_1 := 1 - u_1$, explicitly written as
$$
\Ov u_1 (\alpha) := \begin{cases} 1& \text{if}\; \alpha =0\\ 0 &\text{otherwise} \end{cases}
$$
With the usual conventions the Table \ref{tbl:Ex-LNG-a0} is translated into the Boolean form:
%---
\begin{equation}
\label{tbl:LNG-device-01}
%---
      \begin{tabular}{||c|c||c|c|c|c||c||}
         \hline
         $x_1$ & $x_2$ & $x_3$ & $\longmapsto$ &
                 $x'_1$ & $x'_2$ & $x'_3$ \\
         \hline\hline
         \mbox{\rule[0cm]{0cm}{2.5ex}
         $0$}& $0$ & $1$ & & $0$ & $0$ & $1$\\
         $0$ & $1$ & $1$ & & $0$ & $1$ & $0$ \\
         $1$ & $0$ & $1$ & & $1$ & $0$ & $0$ \\
         \mbox{\rule[0cm]{0cm}{1.5ex}$1$}
             & $1$ & $1$ & & $1$ & $1$ & $0$ \\
         \hline
      %\caption{Andamento fisico del dispositivo LNG quando la lamella \e inizialmente in posizione verticale}
      \end{tabular}
%---
\end{equation}
That is, the fixed bit $x_3=1$ is the ancilla input, whereas $x'_3=\neg(x_1\lor x_2)$ is the required \textsc{Nor} connective produced by the LNG device under the new normalization function. But this is just the same situation described by the self-reversible CL gate depicted in Fig.\til \ref{fig:bi-CLV-NOR} corresponding to the transitions $(a,b,1)\xrightarrow{CL} (a,b,\neg(a\lor b)) \xrightarrow{CL} (a,b,1)$ discussed in section \ref{sec:3-3CL-LNG}.
Therefore we can state the following

\begin{description}
\item[3rd Conclusion]\emph{
The LNG device under the assumptions of the initial cantilever angle $\alpha_i=0$ and the normalized function on angles $\Ov u_1$ is a concrete realization of the CL 3in/3out self-reversible gate with the third input fixed to 1, producing in the third output the connective \textsc{Nor}.}
\end{description}
%-----------------
\subsection{Fourth normalization function: the Toffoli case for generating \textsc{Nand}}
%---------------------------------------
Analogously to the previous case, one can apply to the LNG device described by the Table \ref{tbl:Ex-LNG-a0} the conventional normalization function $\Ov u_2 := 1 - u_2$, explicitly written as
$$
\Ov u_2 (\alpha) := \begin{cases} 1& \text{if}\; 0\le\alpha \le 1\\ 0 &\text{otherwise} \end{cases}
$$
In this case, with the usual conventions the Table \ref{tbl:Ex-LNG-a0} is translated into the Boolean form:
%---
\begin{equation}
\label{tbl:LNG-device-11}
%---
      \begin{tabular}{||c|c||c|c|c|c||c||}
         \hline
         $x_1$ & $x_2$ & $x_3$ & $\longmapsto$ &
                 $x'_1$ & $x'_2$ & $x'_3$ \\
         \hline\hline
         \mbox{\rule[0cm]{0cm}{2.5ex}
         $0$}& $0$ & $1$ & & $0$ & $0$ & $1$\\
         $0$ & $1$ & $1$ & & $0$ & $1$ & $1$ \\
         $1$ & $0$ & $1$ & & $1$ & $0$ & $1$ \\
         \mbox{\rule[0cm]{0cm}{1.5ex}$1$}
             & $1$ & $1$ & & $1$ & $1$ & $0$ \\
         \hline
      %\caption{Andamento fisico del dispositivo LNG quando la lamella \e inizialmente in posizione verticale}
      \end{tabular}
%---
\end{equation}
That is, fixing the third input with the bit $x_3=1$, the third output $x'_3=\neg(x_1\land x_2)$ is the required \textsc{Nand} connective produced by the LNG device under the new normalization function. But this is just the same situation described by the self-reversible Toffoli gate corresponding to the transitions $(a,b,1)\xrightarrow{T} (a,b,\neg(a\land b)) \xrightarrow{T} (a,b,1)$ discussed at the end of section \ref{sec:3-3T}.
Therefore we can state the following

\begin{description}
\item[4th Conclusion]\emph{
The LNG device under the assumptions of the initial cantilever angle $\alpha_i=0$ and the normalized function on angles $\Ov u_2$ is a concrete realization of the Toffoli 3in/3out self-reversible gate with the third input fixed to 1, producing in the third output the connective \textsc{Nand}.}
\end{description}
%--------------------------------------
\section{The particular case of the LNG realization of the connective \textsc{Xor}}
%---------------------------------
In this section we will take into account the self-reversible realization of \textsc{Xor} logic connective by the LNG device, making use as usual of some suitable normalization function. But first of all let us
note that if we take into account the CL self-reversible gate of section \ref{sec:3-3CL-LNG}, this connective can be realized either fixing the first input to 0 (and this in the LNG device corresponds to fixing the voltage of the first probe) or fixing the second input to 0 (and also in this case it is the voltage of the second probe in the LNG device which must be fixed). The same considerations can be done in the Toffoli case where the fixed either first or second input must be 1. This gives rise to a problem as we discuss now.

Let us consider the case of the CL gate for the fixed input $x_1=0$, which produces as the third output the \textsc{Xor} logic connective of the second and third inputs, $x'_3=x_2\oplus x_3$, whose table describing this case is the following:

%---
%\begin{equation}
\begin{table}[h]
      \begin{tabular}{||c||c|c|c|c|c||c||}
         \hline
         $x_1$ & $x_2$ & $x_3$ & $\xrightarrow{G_{CL}}$ &
                 $x'_1$ & $x'_2$ & $x'_3$ \\
         \hline\hline
         \mbox{\rule[0cm]{0cm}{2.5ex}
         $0$}& $0$ & $0$ & & $0$ & $0$ & $0$\\
         $0$ & $1$ & $0$ & & $0$ & $1$ & $1$ \\
         $0$ & $0$ & $1$ & & $0$ & $0$ & $1$ \\
         \mbox{\rule[0cm]{0cm}{1.5ex}$0$}
             & $1$ & $1$ & & $0$ & $1$ & $0$ \\
         \hline
      \end{tabular}
      \vspace{0.3cm}
      \caption{The \textsc{Xor} connective realization by the CL gate as the third output $x'_3=x_2\oplus x_3$ for the first line fixed to $x_1=x'_1=0$}
\label{tab:CL-a=0}
\end{table}
%---
% \end{equation}
%---

But trying to implement the CL gate inputs $x_1 x_2 x_3$ into the LNG device, applying the usual physical behaviors described by (Exp1)--(Exp6) in section \ref{sec:Or-LNG-CLV} and making use of the normalization function of subsection \ref{ssec:CL-norm-Or} relative to the CL case, one obtains the following two tables (the left physical table and the corresponding Boolean one at the right):
%---
\begin{equation}
\label{tbl:CLV-gate-10}
%---
      \begin{tabular}{|c||c|c|c||c||c|c|}
         \hline
         $P_1$ & $P_2$ & $u_1(\alpha_i)$ & $\longmapsto$ &
                 $P'_1$ & $P'_2$ & $u_1(\alpha_o)$ \\
         \hline\hline
         \mbox{\rule[0cm]{0cm}{2.5ex}
         $D$}& $D$ & $0$ & & $D$ & $D$ & $0$\\
         %---$0$ & $0$ & $1$ & & $0$ & $0$ & $1$ \\
         $D$ & $A$ & $0$ & & $D$ & $A$ & $1$ \\
         %---$0$ & $1$ & $1$ & & $0$ & $1$ & $0$ \\
         %---\hline
         $D$ & $D$ & $1$ & & $D$ & $D$ & $0$ \\
         %---$1$ & $0$ & $1$ & & $1$ & $0$ & $0$ \\
         \mbox{\rule[0cm]{0cm}{1.5ex}$D$}
             & $A$ & $1$ & & $D$ & $A$ & $1$ \\
         %---    & $1$ & $1$ & & $1$ & $1$ & $0$\\
         \hline
      %\caption{Andamento fisico del dispositivo LNG quando la lamella \e inizialmente in posizione verticale}
      \end{tabular}
%---
\qquad
%---
      \begin{tabular}{|c||c|c|c||c||c|c|}
         \hline
         $x_1$ & $x_2$ & $x_3$ & $\longmapsto$ &
                 $x'_1$ & $x'_2$ & $x'_3$ \\
         \hline\hline
         \mbox{\rule[0cm]{0cm}{2.5ex}
         $0$}& $0$ & $0$ & & $0$ & $0$ & $0$\\
         %---$0$ & $0$ & $1$ & & $0$ & $0$ & $1$ \\
         $0$ & $1$ & $0$ & & $0$ & $1$ & $1$ \\
         %---$0$ & $1$ & $1$ & & $0$ & $1$ & $0$ \\
         %---\hline
         $0$ & $0$ & $1$ & & $0$ & $0$ & $0$ \\
         %---$1$ & $0$ & $1$ & & $1$ & $0$ & $0$ \\
         \mbox{\rule[0cm]{0cm}{1.5ex}$0$}
             & $1$ & $1$ & & $0$ & $1$ & $1$ \\
         %---    & $1$ & $1$ & & $1$ & $1$ & $0$\\
         \hline
      \end{tabular}
%---
 \end{equation}
As immediate comparison, the Boolean LNG $x'_3$ of the table \eqref{tbl:CLV-gate-10} at the right has nothing to do with $x_2\oplus x_3$ obtained in the Table \ref{tab:CL-a=0} describing the CL case.
In order to overcome this drawback in the present section we propose two solutions.
%-------------------------
\subsection{A first solution for the \textsc{Xor} generation by the LNG device}
%----------------------------------------
The first solution regards what happens if we consider $x'_3:=|u_1(\alpha_i)-u_1(\alpha_o)|$ instead of $u_1(\alpha_o)$.
Let's see if this choice has some interest or possible physical realization.
We will have two cases, each with its two subcases:
 %---
 \begin{description}
 \item[{$|u_1(\alpha_i)-u_1(\alpha_o)|=0$}]
$ $
% corrispondente ai due soli casi possibili:
  \begin{enumerate}
  \item[00,] i.e., if the tip is initially vertical, it remains vertical;
  \item[11,] i.e., if the tip is deflated by $\alpha_i\neq 0$, it remains deflected, although the two angles might be different $\alpha_i\neq \alpha_o$.
  \end{enumerate}
 %---
  \item[{$|u_1(\alpha_i)-u_1(\alpha_o)|=1$}]
  $ $
%  corrispondente ai due soli casi possibili:
  \begin{enumerate}
  \item[01,] i.e., if the tip is initially vertical, after processing it is deflated;
  \item[10,] i.e., if the tip is deflated by $\alpha_i\neq 0$, eventually, it returns to the vertical position.
 \end{enumerate}
 \end{description}

These are, at least in principle, all physically observable, but one has to change (or rather, complete) the left LNG table \eqref{tbl:CLV-gate-10} as follows:
%Queste situazioni sono, almeno in linea di principio, tutte fisicamente osservabili. Ma ci\o comporta una modifica (o meglio, completamento) della tabella  \eqref{tbl:CLV-gate-10} a sinistra nel seguente modo:
%---
\begin{equation}
\label{tbl:CLV-gate-10+}
%---
      \begin{tabular}{|c||c|c|c||c||c|c||c||}
         \hline
         $P_1$ & $P_2$ & $u_1(\alpha_i)$ & $\longmapsto$ &
                 $P'_1$ & $P'_2$ & $u_1(\alpha_o)$ & $|u_1(\alpha_i) -u_1(\alpha_o)|$ \\
         \hline\hline
         \mbox{\rule[0cm]{0cm}{2.5ex}
         $D$}& $D$ & $0$ & & $D$ & $D$ & $0$ & 0\\
         %---$0$ & $0$ & $1$ & & $0$ & $0$ & $1$ \\
         $D$ & $A$ & $0$ & & $D$ & $A$ & $1$ & 1\\
         %---$0$ & $1$ & $1$ & & $0$ & $1$ & $0$ \\
         %---\hline
         $D$ & $D$ & $1$ & & $D$ & $D$ & $0$ & 1\\
         %---$1$ & $0$ & $1$ & & $1$ & $0$ & $0$ \\
         \mbox{\rule[0cm]{0cm}{1.5ex}$D$}
             & $A$ & $1$ & & $D$ & $A$ & $1$ & 0\\
         %---    & $1$ & $1$ & & $1$ & $1$ & $0$\\
         \hline
      %\caption{Andamento fisico del dispositivo LNG quando la lamella \e inizialmente in posizione verticale}
      \end{tabular}
\end{equation}
%---
where, setting $x'_3 = |u_1(\alpha_i) -u_1(\alpha_o)|$, one obtains the following table where the third output is the \textsc{Xor} of the second and third input, while the first and second lines retain their value unchanged:
%---
\begin{equation*}
      \begin{tabular}{|c||c|c|c||c|c||c||}
         \hline
         $x_1$ & $x_2$ & $x_3$ & $\longmapsto$ &
                 $x'_1$ & $x'_2$ & $x'_3$ \\
         \hline\hline
         \mbox{\rule[0cm]{0cm}{2.5ex}
         $0$}& $0$ & $0$ & & $0$ & $0$ & $0$\\
         %---$0$ & $0$ & $1$ & & $0$ & $0$ & $1$ \\
         $0$ & $1$ & $0$ & & $0$ & $1$ & $1$ \\
         %---$0$ & $1$ & $1$ & & $0$ & $1$ & $0$ \\
         %---\hline
         $0$ & $0$ & $1$ & & $0$ & $0$ & $1$ \\
         %---$1$ & $0$ & $1$ & & $1$ & $0$ & $0$ \\
         \mbox{\rule[0cm]{0cm}{1.5ex}$0$}
             & $1$ & $1$ & & $0$ & $1$ & $0$ \\
         %---    & $1$ & $1$ & & $1$ & $1$ & $0$\\
         \hline
      \end{tabular}
%---
 \end{equation*}

Note that even in this case one has to do with the realization of the self-reversible \textsc{Xor} (each output is generated by a single input) as the third output in a 3in/3out gate.
%-----------------------
\subsection{Coherence of the CL through LNG with the present approach}
%-----------------------
The question arises if what is obtained in the section \ref{ssec:CL-norm-Or} is still valid when the left table \eqref{tbl:CLV-gate-30}, in which the only results of $u_1(\alpha_0)$ are considered, is completed with a further final column relative to the outputs $|u_1(\alpha_i)-u_1(\alpha_o)|$.  This leas to the following table:

\begin{equation}
\label{tbl:CLV-gate-300}
%---
      \begin{tabular}{|c|c|c||c|c|c|c||c||}
         \hline
         $P_1$ & $P_2$ & $u_1(\alpha_i)$ & $\longmapsto$ &
                 $P'_1$ & $P'_2$ & $u_1(\alpha_o)$ & $|u_1(\alpha_i)-u_1(\alpha_o)|$\\
         \hline\hline
         \mbox{\rule[0cm]{0cm}{2.5ex}
         $D$}& $D$ & $0$ & & $D$ & $D$ & $0$ & $0$\\
         $D$ & $A$ & $0$ & & $D$ & $A$ & $1$ & $1$\\
         $A$ & $D$ & $0$ & & $A$ & $D$ & $1$ & $1$\\
         \mbox{\rule[0cm]{0cm}{1.5ex}$A$ }
             & $A$ & $0$ & & $A$ & $A$ & $1$ & $1$\\
         \hline
      %\caption{Andamento fisico del dispositivo LNG quando la lamella \e inizialmente in posizione verticale}
      \end{tabular}
\end{equation}
%---
Obviously, both $\hat{x_3} = u_1(\alpha_o)$ and $x'_3 = |u_1(\alpha_i) - u_1(\alpha_o)|$ give always the same value confirming what has been achieved in section \ref{ssec:CL-norm-Or}, where no reference has been made to this fourth output.
%-------------------------
\section{A second solution for the \textsc{Xor} generation by the LNG device and the induced \textsc{NXor} connective}
%----------------------------------------
Let us now consider the second solution for generating the \textsc{Xor} connective consisting in introducing a third self-reversible 3in/3out gate, besides the previously treated CL and Toffoli ones, whose function representation is given by
\begin{equation}
\label{eq:X-fun}
\text{X Gate} =
\begin{cases}
x'_1 = x_1\\
x'_2 = x_2 \\
x'_3 = (x_1\oplus x_2)\oplus  x_3
\end{cases}
\end{equation}

The corresponding tabular representation is given by the Table \ref{tbl:X-gate}.
%---

%\begin{equation}
\begin{table}[h]
      \begin{tabular}{|c|c||c||c||c|c||c||}
         \hline
         $x_1$ & $x_2$ & $x_3$ & $\xrightarrow{G_X}$ &
                 $x'_1$ & $x'_2$ & $x'_3$ \\
         \hline\hline
         \mbox{\rule[0cm]{0cm}{2.5ex}
         $0$}& $0$ & $0$ & & $0$ & $0$ & $0$\\
         $0$ & $0$ & $1$ & & $0$ & $0$ & $1$ \\
         $0$ & $1$ & $0$ & & $0$ & $1$ & $1$ \\
         $0$ & $1$ & $1$ & & $0$ & $1$ & $0$ \\
         \hline
         $1$ & $0$ & $0$ & & $1$ & $0$ & $1$ \\
         $1$ & $0$ & $1$ & & $1$ & $0$ & $0$ \\
         $1$ & $1$ & $0$ & & $1$ & $1$ & $0$ \\
         \mbox{\rule[0cm]{0cm}{1.5ex}$1$}
             & $1$ & $1$ & & $1$ & $1$ & $1$\\
         \hline
      \end{tabular}
      \vspace{0.3cm}
      \caption{Tabular representation of the 3in/3out self-reversible X-gate}
\label{tbl:X-gate}
\end{table}
%\end{equation}
%\newpage

This 3in/3out gate is trivially self-reversible: $G_X\circ G_X= id$. Moreover, from the Table \ref{tbl:X-gate}, choosing as usual some input as fixed ancilla, the following possible cases follow:

 \begin{align*}
 (0,a,b)&\to (0,a,a\oplus b) & (a,0,b)&\to (a,0,a\oplus b)   & (a,b,0)&\to (a, b,a\oplus b)\\
 (1,a,b)&\to (1,a,\neg(a\oplus b))    & (a,1,b)&\to (a,1,\neg(a\oplus b))      & (a,b,1)&\to (a, b,\neg(a\oplus b))
 \end{align*}
%---
corresponding to the realization of the logic connectives \textsc{Xor} $\oplus$ and \textsc{NXor} $\neg\oplus$. From these results we obtain the following realizations of the \textsc{Fan}\textsc{Out} gate and the negation connective \textsc{Not}, respectively,

 \begin{align*}
 (0,a,0)&\to (0,a,a) & (a,0,0)&\to (a,0,a)   &   %---(a,0,0)&\to (a, 0,a)
 \\
 (1,a,0)&\to (1,a,\neg(a))    & (a,1,0)&\to (a,1,\neg(a))      & (a,0,1)&\to (a, 0,\neg(a))
 \end{align*}
%---

Also in this X-gate case we have a controlled-controlled behaviour in the sense that if both the two control lines $x_1$ and $x_2$ are equal, then to the third target line it acts the identity, while if the two control lines $x_1$ and $x_2$ are different, then to the third target line it acts the negation:
\begin{equation*}
(a,b,c)\xrightarrow{a=b}(a,b,c)\qquad (a,b,c)\xrightarrow{a\neq b}(a,b,\neg c)
\end{equation*}

Now, let us give the full partial table from Table \ref{tbl:X-gate} corresponding to the generation as third output of the \textsc{Xor} connective when the third input is fixed as ancilla to the bit 0:

%---
\begin{table}[h]
   %\begin{equation}
      \begin{tabular}{|c|c||c||c|c|c||c||}
         \hline
         $x_1$ & $x_2$ & $x_3$ & $\xrightarrow{G_X}$ &
                 $x'_1$ & $x'_2$ & $x'_3$ \\
         \hline\hline
         \mbox{\rule[0cm]{0cm}{2.5ex}
         $0$}& $0$ & $0$ & & $0$ & $0$ & $0$\\
         $0$ & $1$ & $0$ & & $0$ & $1$ & $1$ \\
         $1$ & $0$ & $0$ & & $1$ & $0$ & $1$ \\
         \mbox{\rule[0cm]{0cm}{1.5ex}$1$}
             & $1$ & $0$ & & $1$ & $1$ & $0$ \\
         \hline
      \end{tabular}
      \vspace{0.3cm}
      \caption{Generation of the \textsc{Xor} connective as third output form the self-reversible 3in/3out X-gate $G_X$, when the third input is fixed to 0: $x'_3=x_1\oplus x_2$}
\label{tab:X-c=0}
   %\end{equation}
%---
 \end{table}
%---
\newpage

Now, taking into account the experimental behavior of the LNG device described by Table \ref{tbl:Ex-LNG-a0} under the condition of initial cantilever angle $\alpha_i=0$ we will try to realize the \textsc{Xor} connective of the above Table \ref{tab:X-c=0} obtained from the 3in/3out self-reversible X-gate. We need at this purpose to consider a peculiar normalization function assigning Boolean values 0 and 1 to the possible cantilever angles. The required normalization function is the following one:
\begin{equation}\label{eq:u3}
u_3(\alpha) := \begin{cases}0 & \text{if}\;\; \alpha=0\\
                            1 & \text{if}\;\; 0<\alpha\le 1\\
                            0 & \text{if}\;\; \alpha > 1 \end{cases}
\end{equation}
obtaining from the Table \ref{tbl:Ex-LNG-a0} the following \virg{normalized} two tables: the one on the left is the physical behavior of the LNG device with the normalization of the cantilever angles and the one on the right corresponding to its Boolean version under the usual conventions $D=0$ and $A=1$:
\begin{equation}
\label{tbl:LNG-gate-X}
%---
      \begin{tabular}{|c|c||c||c|c|c||c||}
         \hline
         $P_1$ & $P_2$ & $u_3(\alpha_i)$ & $\longmapsto$ &
                 $P'_1$ & $P'_2$ & $u_3(\alpha_o)$ \\
         \hline\hline
         \mbox{\rule[0cm]{0cm}{2.5ex}
         $D$}& $D$ & $0$ & & $D$ & $D$ & $0$\\
         $D$ & $A$ & $0$ & & $D$ & $A$ & $1$ \\
         $A$ & $D$ & $0$ & & $A$ & $D$ & $1$ \\
         \mbox{\rule[0cm]{0cm}{1.5ex}$A$}
             & $A$ & $0$ & & $A$ & $A$ & $0$ \\
         \hline
      %\caption{Andamento fisico del dispositivo LNG quando la lamella \e inizialmente in posizione verticale}
      \end{tabular}
%---
\qquad
%---
      \begin{tabular}{|c|c||c||c|c|c||c||}
         \hline
         $x_1$ & $x_2$ & $x_3$ & $\longmapsto$ &
                 $x'_1$ & $x'_2$ & $x'_3$ \\
         \hline\hline
         \mbox{\rule[0cm]{0cm}{2.5ex}}
         $0$ & $0$ & $0$ & & $0$ & $0$ & $0$\\
         $0$ & $1$ & $0$ & & $0$ & $1$ & $1$ \\
         $1$ & $0$ & $0$ & & $1$ & $0$ & $1$ \\
         \mbox{\rule[0cm]{0cm}{1.5ex}$1$}
             & $1$ & $0$ & & $1$ & $1$ & $0$ \\
         \hline
      \end{tabular}
%---
 \end{equation}
 where the Boolean version of the LNG device behavior presented by the table at the right coincides just with the Table \ref{tab:X-c=0} obtained by the 3in/3out self-reversible X-gate under the assumption $x_3=0$.
 This leads to the following:

\begin{description}
\item[5th Conclusion]\emph{
The LNG device under the assumptions of the initial cantilever angle $\alpha_i=0$ and the normalized function on angles $u_3$ is a concrete realization of the 3in/3out self-reversible X-gate with the third input fixed to 0, producing in the third output the connective \textsc{Xor}.}
\end{description}
%--------------------
\subsection{The \textsc{NXor} connective generation by LNG}
%------------------
 Let us note that if in the Table \ref{tbl:X-gate}, giving the tabular representation of the X-gate, the third input is fixed as ancilla to the bit 1, then one obtains the partial table corresponding to the generation as third output of the \textsc{NXor} connective

%---
\begin{table}[h]
   %\begin{equation}
      \begin{tabular}{|c|c||c||c|c|c||c||}
         \hline
         $x_1$ & $x_2$ & $x_3$ & $\xrightarrow{G_X}$ &
                 $x'_1$ & $x'_2$ & $x'_3$ \\
         \hline\hline
         \mbox{\rule[0cm]{0cm}{2.5ex}
         $0$}& $0$ & $1$ & & $0$ & $0$ & $1$\\
         $0$ & $1$ & $1$ & & $0$ & $1$ & $0$ \\
         $1$ & $0$ & $1$ & & $1$ & $0$ & $0$ \\
         \mbox{\rule[0cm]{0cm}{1.5ex}$1$}
             & $1$ & $1$ & & $1$ & $1$ & $1$ \\
         \hline
      \end{tabular}
      \vspace{0.3cm}
      \caption{Generation of the \textsc{NXor} connective form the self-reversible 3in/3out X-gate as third output, when the third input $x_3$ is fixed to 1: $x'_3=\neg(x_1\oplus x_2)$}
\label{tab:NX-c=0}
   %\end{equation}
%---
 \end{table}
%---

Now, if instead of the angle normalization function \eqref{eq:u3} one considers its \virg{negation} $\Ov u_3:= 1 - u_3$ explicitly defined by the rules
\begin{equation}\label{eq:u3}
\Ov u_3(\alpha) := \begin{cases}1 & \text{if}\;\; \alpha=0\\
                            0 & \text{if}\;\; 0<\alpha\le 1\\
                            1 & \text{if}\;\; \alpha > 1 \end{cases}
\end{equation}
the Table \ref{tbl:Ex-LNG-a0} describing the LNG physical behavior becomes (with the table at the right giving its Boolean version)
\begin{equation}
\label{tbl:LNG-gate-NX}
%---
      \begin{tabular}{|c|c||c||c|c|c||c||}
         \hline
         $P_1$ & $P_2$ & $\Ov u_3(\alpha_i)$ & $\longmapsto$ &
                 $P'_1$ & $P'_2$ & $\Ov u_3(\alpha_o)$ \\
         \hline\hline
         \mbox{\rule[0cm]{0cm}{2.5ex}
         $D$}& $D$ & $1$ & & $D$ & $D$ & $1$\\
         $D$ & $A$ & $1$ & & $D$ & $A$ & $0$ \\
         $A$ & $D$ & $1$ & & $A$ & $D$ & $0$ \\
         \mbox{\rule[0cm]{0cm}{1.5ex}$A$}
             & $A$ & $1$ & & $A$ & $A$ & $1$ \\
         \hline
      %\caption{Andamento fisico del dispositivo LNG quando la lamella \e inizialmente in posizione verticale}
      \end{tabular}
%---
\qquad
%---
      \begin{tabular}{|c|c||c||c|c|c||c||}
         \hline
         $x_1$ & $x_2$ & $x_3$ & $\longmapsto$ &
                 $x'_1$ & $x'_2$ & $x'_3$ \\
         \hline\hline
         \mbox{\rule[0cm]{0cm}{2.5ex}}
         $0$ & $0$ & $1$ & & $0$ & $0$ & $1$\\
         $0$ & $1$ & $1$ & & $0$ & $1$ & $0$ \\
         $1$ & $0$ & $1$ & & $1$ & $0$ & $0$ \\
         \mbox{\rule[0cm]{0cm}{1.5ex}$1$}
             & $1$ & $1$ & & $1$ & $1$ & $1$ \\
         \hline
      \end{tabular}
%---
 \end{equation}
 where the output $x'_3 = \neg(x_1\oplus x_2)$ of the table at the right is the \textsc{NXor} connective of the inputs $x_1$ and $x_2$, negation of the \textsc{Xor} connective described by the tables \eqref{tbl:LNG-gate-X}, leading to the further

\begin{description}
\item[6th Conclusion]\emph{
The LNG device under the assumptions of the initial cantilever angle $\alpha_i=0$ and the normalized function on angles $\Ov u_3$ is a concrete realization of the 3in/3out self-reversible X-gate with the third input fixed to 1, producing in the third output the connective \textsc{NXor}.}
\end{description}
\section{Conclusions}
%------------------
In this paper we discussed an experimental device proposed by L\'opez-Su\'arez et al.\til in \cite{LNG16}, whose essential description has been synthesized by us in section \ref{sec:LNG-Or}, especially their conclusion that it realizes by a micro-electromechanical cantilever the classical irreversible \textsc{Or} logic gate, operating with energy below the  expected limit stated in literature as Landauer principle.

Our analysis of the LNG experimental device, performed first in section \ref{sec:1st-LNG} and then more deeply treated in section \ref{sec:3-3CL-LNG}, arrives to the conclusion that the LNG experimental device can be described as the realization of a 3in/3out self-reversible gate whose \textsc{Or} logic connective is obtained with the usual procedure of fixing the third input as ancilla of logic value 0, considering the first two outputs as garbage and obtaining in this way as third output the required connective as shown in Fig.\til \ref{fig:OR-CL} and \ref{fig:bi-CLV-OR}.
This is obtained as a Cattaneo-Leporini (CL) 3in/3out self-reversible gate if one adopts a normalization of the experimental angles by a suitable function as discussed in subsection \ref{ssec:CL-norm-Or}. Owing to the self-reversibility of this gate there is no contradiction with the results of arbitrary small dissipation of energy, i.e., well below $k_B\ T$, experimentally obtained by the LNG device. On the other hand, and on the basis of the Toffoli (T) 3in/3out self-reversible gate, making use of another suitable angle normalization function as discussed in subsection \ref{ssec:T-norm-And}, from the LNG device it is possible to obtain the \textsc{And} logic gate with the usual procedure \virg{ancilla-inputs-garbage-output} and so, also in this case, with arbitrary small energy dissipation without any contradiction with the Landauer principle.

This procedure consisting of the results of the given LNG device with initial cantilever angle equal to 0 and suitable functions normalizing the cantilever angles, can be extended to obtain  also the logic connectives \textsc{Nor}, \textsc{Nand}, \textsc{Xor} and \textsc{NXor} in a self-reversible way.

This leads to consider the pair formed by the LNG device plus a \virg{\emph{memory}} containing all the necessary normalization functions $\parg{u_1,u_2,\ldots,u_j,\ldots,u_6}$ as a \emph{universal logic machine} in the sense that based on the LNG device, by the input of a suitable angle normalization function $u_j$, it is possible to obtain one of logic connective from the collection
$$
LC_{LNG} = \parg{\textsc{Or},\textsc{And},\textsc{XOr},\textsc{NOr},\textsc{NAnd},\textsc{NXor}}
$$
This universal logic machine is schematized in the below figure

\begin{figure}[ht]
\begin{center}
   \includegraphics[width=8cm]{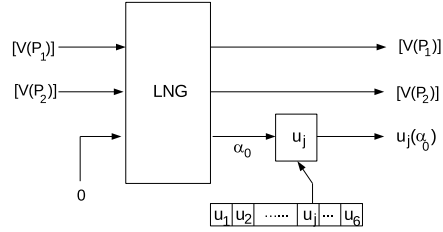}
\end{center}
   \vspace{-1.0cm}
      \caption{Schemata of the LNG-Machine with the memory of the cantilever angle normalized functions}
      \label{fig:LNG-Mac}
\end{figure}

Let us recall that all these logic connectives generated by the LNG device are obtained under the assumption introduced in section \ref{sec:Or-LNG-CLV} that the two output cantilever angles $\hat\alpha_1$ and $\tilde{\alpha}_1$, as experimentally very similar, are considered as equal (indistinguishable) between them. Let us now suppose, as discussed in section \ref{sec:1st-LNG}, that after some technological development these two angles can be detected as different. In this case we must modify the Table \ref{tbl:Ex-LNG-a0} in order to take into account this difference in the following way:

\begin{table}[h]
%---
      \begin{tabular}{|c|c||c||c|c|c||c||}
         \hline
         $P_1$ & $P_2$ & $\alpha_i$ & $\longmapsto$ &
                 $P'_1$ & $P'_2$ & $\alpha_o$ \\
         \hline\hline
         \mbox{\rule[0cm]{0cm}{2.5ex}
         $D$}& $D$ & $0$ & & $D$ & $D$ & $0$\\
         $D$ & $A$ & $0$ & & $D$ & $A$ & $\hat\alpha_1$ \\
         $A$ & $D$ & $0$ & & $A$ & $D$ & $\tilde\alpha_1$ \\
         \mbox{\rule[0cm]{0cm}{1.5ex}$A$}
             & $A$ & $0$ & & $A$ & $A$ & $\alpha_2$ \\
         \hline
      \end{tabular}
      \vspace{0.3cm}
      \caption{Physical behavior of the LNG device when the cantilever is in the initial vertical position $\alpha_i=0$ and with the final angles $\alpha_o$ ordering $0<\hat\alpha_1<\tilde{\alpha}_1< 1 <\alpha_2$, with $\hat{\alpha}_1\neq\tilde{\alpha}_1$}
\label{tbl:Ex-LNG-dif}
\end{table}

Let us now introduce the further cantilever angle normalization function whose simplified version is the following one
$$
u_4(\alpha) :=\begin{cases}
               1 & \text{if}\;\;\alpha=0\\
               1 & \text{if}\;\;\alpha=\hat{\alpha}_1\\
               0 & \text{if}\;\;\alpha=\tilde{\alpha}_1\\
               1 & \text{if}\;\;\alpha=\alpha_2\\
              \end{cases}
$$

Using this normalization function, the above Table \ref{tbl:Ex-LNG-dif} assumes the Boolean form under the usual conventions $D=0$ and $A=1$:

\begin{table}[h]
%---
      \begin{tabular}{|c|c||c||c|c|c||c||}
         \hline
         $x_1$ & $x_2$ & $x_3=u_4(\alpha_i)$ & $\longmapsto$ &
                 $x'_1$ & $x'_2$ & $x'_3=u_4(\alpha_o)$ \\
         \hline\hline
         \mbox{\rule[0cm]{0cm}{2.5ex}
         $0$}& $0$ & $1$ & & $0$ & $0$ & $1$\\
         $0$ & $1$ & $1$ & & $0$ & $1$ & $1$ \\
         $1$ & $0$ & $1$ & & $1$ & $0$ & $0$ \\
         \mbox{\rule[0cm]{0cm}{1.5ex}$1$}
             & $1$ & $1$ & & $1$ & $1$ & $1$ \\
         \hline
      \end{tabular}
      \vspace{0.3cm}
      \caption{Boolean form under the normalization function $u_4$ of the above LNG experimental behavior given by Table \ref{tbl:Ex-LNG-dif}}
\label{tbl:Ex-LNG-dif-Bo}
\end{table}

where the third output $x'_3 = \neg x_1 \lor x_2 = x_1 \to x_2$ is the \emph{implication} connective of the first two inputs $x_1$ and $x_2$.
But if one consider the self-reversible 3in/3out gate whose tabular representation is the following one

\begin{table}[h]
      \begin{tabular}{|c|c||c||c||c|c||c||}
         \hline
         $x_1$ & $x_2$ & $x_3$ & $\xrightarrow{G_I}$ &
                 $x'_1$ & $x'_2$ & $x'_3$ \\
         \hline\hline
         \mbox{\rule[0cm]{0cm}{2.5ex}
         $0$}& $0$ & $0$ & & $0$ & $0$ & $0$\\
         $0$ & $0$ & $1$ & & $0$ & $0$ & $1$ \\
         $0$ & $1$ & $0$ & & $0$ & $1$ & $0$ \\
         $0$ & $1$ & $1$ & & $0$ & $1$ & $1$ \\
         \hline
         $1$ & $0$ & $0$ & & $1$ & $0$ & $1$ \\
         $1$ & $0$ & $1$ & & $1$ & $0$ & $0$ \\
         $1$ & $1$ & $0$ & & $1$ & $1$ & $0$ \\
         \mbox{\rule[0cm]{0cm}{1.5ex}$1$}
             & $1$ & $1$ & & $1$ & $1$ & $1$\\
         \hline
      \end{tabular}
      \vspace{0.3cm}
      \caption{Tabular representation of the 3in/3out self-reversible I-gate}
\label{tbl:I-gate}
\end{table}

the partial table corresponding to the third input $x_3=1$ is just coincident with Table \ref{tbl:Ex-LNG-dif-Bo}, which turns out to be a realization of this self-reversible gate by the LNG device with the normalization $u_4$. So also the implication connective can be realized in a self-reversible way, i.e., with arbitrary small dispersion of energy, by the LNG device when the two angle outputs $\hat{\alpha}_1$ and $\tilde{\alpha}_1$ can be detected as different between them.

In conclusion, also if the interpretation of the LNG device as a generator in an irreversible way of the connective \textsc{Or} in an experimental situation of arbitrary small energy dispersion (contrary to the Landauer principle), is erroneous, the LNG device is a very powerful tool for being the essential component of a universal logic machine able to produce, with a suitable choice of a normalization function collected in the memory, a great number of logic connectives.
%================================
\begin{comment}
\begin{table}[h]
      \begin{tabular}{|c|c||c||c||c|c||c||}
         \hline
         $x_1$ & $x_2$ & $x_3$ & $\xrightarrow{G_I}$ &
                 $x'_1$ & $x'_2$ & $x'_3$ \\
         \hline\hline
         \mbox{\rule[0cm]{0cm}{2.5ex}
         $0$ & $0$ & $1$ & & $0$ & $0$ & $1$ \\
         $0$ & $1$ & $1$ & & $0$ & $1$ & $0$ \\
         $1$ & $0$ & $1$ & & $1$ & $0$ & $1$ \\
         \mbox{\rule[0cm]{0cm}{1.5ex}$1$}
             & $1$ & $1$ & & $1$ & $1$ & $0$\\
         \hline
      \end{tabular}
      \vspace{0.3cm}
      \caption{Tabular representation of the 3in/3out self-reversible I-gate}
\label{tbl:I-gate-1}
\end{table}
\end{comment}
%=========================================

%---------------------------------------------------
%\bibliographystyle{amsalpha}    %---{amsplain} fornisce nomi, invece di numeri %--{mysiam}
                                %\bibliography{c:/usr/bib-qm/ref-qcom-temp}
%\bibliography{c:/usr/bib-qm/ref-qcomp}

\begin{thebibliography}{LSNG16}

\bibitem[Fey96]{Fe96}
R.~P. Feynman, \emph{Lectures in computation}, Penguin Books, London, 1996,
  Edited by A.J.G. Hey and R.W. Allen.

\bibitem[FT82]{FT82}
E.~Fredkin and T.~Toffoli, \emph{Conservative logic}, Int.~J.~Theor.~Phys.
  \textbf{21} (1982), 219--253.

\bibitem[KTR12]{KTR12}
S.~Kotiyal, H.~Thapliyal, and N.~Ranganathan, \emph{{Mach}-{Z}ehnder
  interferometer based all optical reversible nor gates}, IEEE Computer Society
  Annual Symposium on {VLSI} (2012), 207--212.

\bibitem[LSNG16]{LNG16}
M.~L\'opez-Su\'arez, I.~Neri, and L.~Gammaitoni, \emph{Sub-$k_{B} {T}$
  micro-electromechanical irreversible logic gate}, Nature Communications
  \textbf{7} (1916), 1--6.

\bibitem[NC00]{NC00}
M.~A. Nielsen and I.~L. Chuang, \emph{Quantum computation and quantum
  information}, Cambridge Univ.~Press, Cambridge, 2000.

\bibitem[Per85]{Pe85}
A.~Peres, \emph{Reversible logic and quantum computers}, Physical Review
  \textbf{A 32} (1985), 3266--3276.

\bibitem[Tof80]{To80}
T.~Toffoli, \emph{Reversible computing}, Laboratory for Computer Science
  MIT/LCS/TM--151, Massachussetts Institute of Technology, 1980.

\end{thebibliography}
%\begin{thebibliography}{99}

%=====================================================
\end{document}